\documentclass[12pt]{article}

\usepackage{amsmath,amssymb}
\usepackage{graphicx}% Include figure files
\usepackage{caption}
\usepackage{color}% Include colors for document elements
\usepackage{dcolumn}% Align table columns on decimal point
\usepackage{bm}% bold math
\usepackage[numbers,super,comma,sort&compress]{natbib}

\definecolor{background-color}{gray}{0.98}

\title{A Study of Entangled Systems in the Many-Body\\Signed Particle Formulation of Quantum Mechanics}
\author{Jean~Michel~Sellier\thanks{IICT, Bulgarian Academy of Sciences, Acad. G.~Bonchev str. 25A, 1113 Sofia, Bulgaria, jeanmichel.sellier@parallel.bas.bg, jeanmichel.sellier@gmail.com}, K.G.~Kapanova\thanks{IICT, Bulgarian Academy of Sciences, Acad. G.~Bonchev str. 25A, 1113 Sofia, Bulgaria}}

\begin{document}

\maketitle

\begin{abstract}
Recently a new formulation of quantum mechanics has been introduced, based on {\sl{signed}} classical field-less particles
interacting with an external field by means of only creation and annihilation events. In this paper, we extend this novel
theory to the case of many-body problems. We show that, when restricted to spatial finite domains and discrete momentum space,
the proposed extended theory is equivalent to the time-dependent many-body Wigner Monte Carlo method. In this new picture,
the treatment of entangled systems comes naturally and, therefore, we apply it to the study of quantum entangled systems.
The latter is represented in terms of two Gaussian wave packets moving in opposite directions. We introduce the presence
of a dissipative background and show how the entanglement is affected by different (controlled) configurations. 
\end{abstract}

\clearpage

%*****************Graphical Table of Contents******************** THIS IS MANDATORY *******************
% Please use the other provided template for writing a graphical ToC. It must not be a part of your main manuscript file. Submit both files separately.

% makes references listed with 1., 2., etc.  
  \makeatletter
  \renewcommand\@biblabel[1]{#1.}
  \makeatother

\bibliographystyle{apsrev}

\renewcommand{\baselinestretch}{1.5}
\normalsize

\clearpage

\section*{\sffamily \Large INTRODUCTION} % Not needed for rapid communications

Quantum entanglement, as one of the most enigmatic and subtle phenomena in nature, has been a subject of
profound interest to scientists from the early development days of quantum mechanics \cite{Einstein},
\cite{Schroedinger_1}, \cite{Schroedinger_2}. Historically, both A.~Einstein and E.~Schr\"{o}dinger
were deeply puzzled by the fact that the existence of this typically quantum phenomenon in the
theory seemed to violate the speed limit at which information can be transmitted. 
%The subsequent development of Bell's inequalities, was as an imperative attempt to provide quantitative measurement of the quantum interaction responsible for the peculiar effects of entanglement \cite{Bell}, \cite{Aspect}. Bell theoretically demonstrated that the principle of locality is incompatible with certain conjectures of quantum mechanics.

The fascinating theoretical consequences of the existence of entangled systems has been investigated
as a potential phenomenon for novel practical applications. Considerable work, assisted by the
technological advances in the latter part of the $20-th$ century, has been extended in both theoretical
and experimental domains to demonstrate the possibility of entanglement with photons \cite{photons},
electrons and atoms \cite{atoms}, molecules \cite{molecules}, as well macroscopic objects such as small
diamonds \cite{diamonds}.
%the existence of non-local effects (albeit a conclusive answer has yet to be presented). For instance, 

From a practical standpoint, recent advances in utilizing entanglement effects in the fields of
quantum information processing and quantum communication have produced promising results, especially
in the growing fields of superdense coding and quantum teleportation \cite{teleportation}. Despite
the remarkable experimental effort devoted to the classification, quantification and utilization
of quantum entanglement, the perception of the unexpected possibilities of entanglement in the design
of novel technological frameworks is still in its infancy. Progress in understanding the physics of
quantum many-body systems has been hindered not only by the enormous complexity of the problems,
but also the formidable computational obstacles.

%Given the new found capacity of entanglement as a new way to understand many physical phenomena,
Recently, a new formulation of quantum mechanics based on the concept of signed particles has been
introduced \cite{SPF} which is able to simulate quantum systems in a time-dependent fashion by means
of ensembles of Newtonian field-less particles. These point-like objects have a position, a momentum,
and a sign, and they allow the time-dependent simulation of complex quantum systems by means of
relatively small computational resources. Notably, the restriction of the formalism to finite domains
and semi-discrete phase-spaces provides a correspondence to the well-known signed particle Wigner
Monte Carlo method. The latter has been thoroughly validated in the past few years \cite{PhysRep},
augmenting it to the case of time-dependent, ab-initio (or first-principle) simulations of quantum
many-body problems \cite{JCP-01}, \cite{JCP-02}, \cite{JCP-03}.

In this work, the signed particle formulation is extended to the investigation of quantum many-body systems.
We then demonstrate that through the restriction to finite domains and semi-discrete phase-spaces this
formalism is equivalent to the many-body Wigner Monte Carlo method. We respectively apply the depicted
novel theory to the study of entangled quantum systems (in particular two-body systems) immersed in a
(controlled) perturbative background. In this paper, we follow the work in \cite{Ferry} which considers a
system of two entangled particles as represented by two Gaussian wave packets moving in opposite directions.
It is established that this corresponds to a quasi-distribution function with an "extra" term representing
the entanglement. The reader is invited to refer to this work for additional details.

The paper is structured as follows. Starting from a conceptualization of the signed particle formulation
of quantum mechanics in the many-body context, the next section introduces how entanglement can be \textbf{viewed}
through the new formalism by means of signed particles only. We establish through
numerical experiments the behavior of entanglement in the presence of an external perturbation. 
Simulations of the same system are performed in the presence of different amounts of noise and 
its effect on the entanglement. We then proceed with a study on the resilience of
quantum entanglement. In particular, we show that by controlling the {\sl{form}} of the
entanglement it is possible to make it stronger or, in other words, less sensitive to external
perturbations. Finally, the new formalism is put into broader framework of possible applications
and future development.

\section*{\sffamily \Large The signed particle formulation for many-body problems}

The following section details how the signed particle formulation of quantum mechanics is extended to
{\sl{many-body}} problems in terms of the three postulates which completely define the theory. The approach
provides advantages from theoretical and numerical perspective. In the former case, it allows for novel
outlooks to the comprehension of quantum systems (further description is provided below). From a numerical point
of view, the theory introduces a straightforward computational implementation of the formalism. The proposed
set of postulates is sufficient to reconstruct the time-dependent evolution of a quantum many-body system
in terms of creation and annihilation events {\sl{only}}. The reader should note that it is feasible to
show that these rules are recovered from a mathematical manipulation and a physical interpretation
of the many-body time-dependent Wigner equation which is equivalent to the time-dependent many-body
Schr\"{o}dinger equation (via the invertible Wigner-Weyl transform). One therefore, applies
the strategy from the single-body formalism \cite{SPF} to the many-body Wigner equation. Importantly,
while the presentation of postulates I and III (below) is in principle preserved with respect to
the single-body formulation, in the current work, they represent, accordingly, point-like particles
evolving in a $(2 \times n \times d)$-dimensional phase-space, thereupon characterizing very different objects.

\bigskip

{\sffamily \Large The postulates}

We henceforth denote by $n=1,2,3,\dots$ the total number of {\sl{physical}} particles (or equivalently {\sl{bodies}})
involved in the system under study, and by $d=1,2,3$ the dimensionality of space.
Consequently, the phase-space coordinates of the $n$-th (virtual) particle are expressed by
$\left( {\bf{x}}^n; {\bf{p}}^n \right) = \left( x_1^n, \dots, x_d^n; p_1^n, \dots, p_d^n  \right)$ 
with a specific configuration of the system provided by the phase-space
point $({\bf{x}}; {\bf{p}}) = \left( {\bf{x}}^1, \dots, {\bf{x}}^n; {\bf{p}}^1, \dots, {\bf{p}}^n \right)$.
Furthermore, in the many-body context a virtual signed particle is understood as a
classical field-less particle where its dynamics are dependent on the following set of equations (in the time interval $[t_0,t]$):
\begin{equation}
 \left( {\bf{x}}^1 (t), \dots, {\bf{x}}^n (t) \right) = \left( {\bf{x}}^1(t_0)+\frac{{\bf{p}}^1(t_0)}{m_1} (t-t_0), \dots, {\bf{x}}^n(t_0)+\frac{{\bf{p}}^n(t_0)}{m_n} (t-t_0) \right),
\end{equation}
with $m_n$ being the mass of the $n$-th body (from here the extension to curved space-time is inconsequential \cite{SPF}).

\bigskip

The postulates now read:

\bigskip

{\sl{{\bf{Postulate I.}} Physical systems can be described by means of (virtual) signed particles defined in the ($2 \times n \times d$-dimensional) phase-space, i.e. provided with a ($n \times d$-dimensional) position ${\bf{x}}$ and a ($n \times d$-dimensional) momentum ${\bf{p}}$ simultaneously, and which carry a sign which can be positive or negative.}}

\bigskip

{\sl{{\bf{Postulate II.}} A signed particle, evolving in a potential $V=V \left( {\bf{x}}^1, \dots, {\bf{x}}^n \right)$, behaves as a
field-less classical point-particle which, during the time interval $dt$, creates a new pair of signed particles
with a probability $\gamma \left( {\bf{x}}^1(t), \dots, {\bf{x}}^n(t) \right) dt$ where
\begin{equation}
 \gamma\left( {\bf{x}}^1, \dots, {\bf{x}}^n \right)=\int_{-\infty}^{+\infty} \mathcal{D}{{\bf{p}}_1}' \dots \int_{-\infty}^{+\infty} \mathcal{D}{{\bf{p}}_n}' V_W^+ \left( {\bf{x}}^1, \dots, {\bf{x}}^n; {{\bf{p}}_1}', \dots, {{\bf{p}}_n}' \right)
\label{momentum_integral}
\end{equation}
is the many-body {\sl{momentum}} integral defined as
\begin{equation}
 \lim_{\Delta {{\bf{p}}_1}' \rightarrow 0^+} \dots \lim_{\Delta {{\bf{p}}_n}' \rightarrow 0^+} \sum_{{{\bf{M}}_1} = -\infty}^{+\infty} \dots \sum_{{{\bf{M}}_n} = -\infty}^{+\infty} V_W^+ \left( {\bf{x}}^1, \dots, {\bf{x}}^n; {\bf{M}}_1 \Delta {{\bf{p}}_1}', \dots, {\bf{M}}_n \Delta {{\bf{p}}_n}' \right),
\label{momentum_integral_definition}
\end{equation}
and $V_W^+ \left( {\bf{x}}^1, \dots, {\bf{x}}^n; {\bf{p}}^1, \dots, {\bf{p}}^n \right)$ is the positive part of
the many-body Wigner kernel \cite{PhysRep}, \cite{JCP-02} and \textbf{$M_1, M_2, \dots, M_{max}$} are integers.
If, at the moment of creation, the parent particle has sign $s$,
position ${\bf{x}} = \left( {\bf{x}}^1, \dots, {\bf{x}}^n \right)$ and momentum ${\bf{p}} = \left( {\bf{p}}^1, \dots, {\bf{p}}^n \right)$,
the new particles are both located in ${\bf{x}}$, have signs $+s$ and $-s$, and momenta ${\bf{p}}+{\bf{p}}'$ and ${\bf{p}}-{\bf{p}}'$ respectively,
with ${\bf{p}}'$ chosen randomly according to the (normalized) probability
$$
\frac{V_W^+ \left( {\bf{x}}^1, \dots, {\bf{x}}^n; {\bf{p}}^1, \dots, {\bf{p}}^n \right)}{\gamma\left({\bf{x}}^1, \dots, {\bf{x}}^n\right)}.
$$}}

\bigskip

{\sl{{\bf{Postulate III.}} Two particles with opposite sign and same phase-space coordinates $\left( {\bf{x}}, {\bf{p}}\right)$ annihilate.}}

\bigskip

{\sffamily \Large Connection with the many-body Wigner Monte Carlo method}

One may obtain a {\sl{finite}} and {\sl{semi-discrete}} phase-space by adapting Postulate II.
Thus, the reformulation of the postulate introduces a momentum step $\Delta p = \frac{\hbar \pi}{L_C}$
(with $L_C$ a given length\footnote{The reader should note that a study on how the parameter $L_C$ affects
the accuracy of the solution has been provided in \cite{Sellier-SA} showing essentially that this variable
can be chosen safely in a wide numerical range.}) in the following manner.

\bigskip

A many-body signed particle with phase-space coordinates
$$
\left( {\bf{x}}^1(t), \dots, {\bf{x}}^n(t); {\bf{M}}^1(t) \Delta p, \dots, {\bf{M}}^n(t) \Delta p\right)
$$
and evolving in a potential $V=V \left( {\bf{x}}^1, \dots, {\bf{x}}^n \right)$, behaves as a
field-less classical point-particle. During the time interval $dt$, the particle creates a new pair of signed particles
with a probability $\gamma \left( {\bf{x}}^1(t), \dots, {\bf{x}}^n(t) \right) dt$
\begin{equation}
 \gamma\left( {\bf{x}}^1, \dots, {\bf{x}}^n \right)=\sum_{{{\bf{M}}_1} = -M_{max}}^{+M_{max}} \dots \sum_{{{\bf{M}}_n} = -M_{max}}^{+M_{max}} V_W^+ \left( {\bf{x}}^1, \dots, {\bf{x}}^n; {\bf{M}}_1 \Delta {{\bf{p}}_1}, \dots, {\bf{M}}_n \Delta {{\bf{p}}_n} \right),
\label{momentum_integral_discrete}
\end{equation}
with the integer $M_{max}$ depicting the {\sl{finiteness}} of the phase-space,
and $V_W^+ \left( {\bf{x}}^1, \dots, {\bf{x}}^n; {\bf{p}}^1, \dots, {\bf{p}}^n \right)$
representing the positive part of the many-body Wigner kernel \cite{PhysRep}, \cite{JCP-02}.

Considering the moment of creation, where the parent particle has sign $s$,
position ${\bf{x}} = \left( {\bf{x}}^1, \dots, {\bf{x}}^n \right)$ and momentum ${\bf{p}} = \left( {\bf{p}}^1, \dots, {\bf{p}}^n \right)$,
the new particles therefore are both located in ${\bf{x}}$, with signs $+s$ and $-s$, and momenta ${\bf{p}}+{\bf{p}}'$ and ${\bf{p}}-{\bf{p}}'$ respectively,
and where ${\bf{p}}'$ is chosen randomly according to the (normalized) probability
$$
\frac{V_W^+ \left( {\bf{x}}^1, \dots, {\bf{x}}^n; {\bf{p}}^1, \dots, {\bf{p}}^n \right)}{\gamma\left({\bf{x}}^1, \dots, {\bf{x}}^n\right)}.
$$

\bigskip

The reader should consider that in this particular case, discernible from
the postulates above, the signed particle formulation corresponds to the
many-body Wigner Monte Carlo method described in \cite{JCP-02}. Indeed,
the adoption of this version of postulate II in the current work (and as
a consequence, the many-body Wigner Monte Carlo method), provides certain
advantages, which require further investigation. Highlighting the
fact that the momentum integral (\ref{momentum_integral}) is not a Riemann
integral, as well as that its convergence is not guaranteed (notably, there
is a significant resemblance with the path integral definition \cite{Feynman}),
several compelling lines in inquiry could be approached. By utilizing the
continuum formulation, an analytical study of the convergence becomes
possible and one could, perhaps, predict that the exact conditions under
which the momentum integral is convergent may provide new insights about
quantum mechanical phenomena. Moreover, one could imagine quantum systems
for which an analytical treatment would be simply impractical in any formulation
but the signed particle one (an example is the applicability of
the most common formulations according to their practical convenience
according to specific context, e.g. the classical limit is trivial in
the Wigner formalism but very perplexing when approached by means of
wave-functions).\footnote{A fact supporting this claim comes from the
development of the path integral formulation. By reading sections 3 and
4 of \cite{Feynman}, it looks most likely that the concept of path integral
was firstly approached by R.~Feynman in a discrete context which was
subsequently extended to a continuum context. It is undeniable that such
approach has brought insights into the mechanics of quantum systems which
would never be possible to reach otherwise. The author firmly suspects
that this could be the case for the signed particle formulation as well.}

\bigskip

{\sffamily \Large The operator $\hat{S}$, scaling and computational complexity}

The many-body formulation (analogous to the single-body case) in accordance
to the uncertainty principle, initiates with an {\sl{ensemble}} of particles
depicting the system in a state of minimal uncertainty (the reader should be
cognizant of the fact that the corresponding quasi-distribution function must
fulfill the conditions in \cite{tatarskii} in order to be physically meaningful).
In the formalism, the definition of an operator $\hat{S}$ constitutes the
creation of a new set of {\sl{three}} signed particles (corresponding to the
single-body case, with the distinction that a bigger phase-space is involved).
Given a signed particle (representing a particular state of the complete system) from the initial ensemble,
at time $t$, with sign $s$, mass $m$ (for the sake of simplicity, we suppose that every involved body in the problem has the same mass),
and phase-space coordinates $\left({\bf{x}}_1, \dots, {\bf{x}}_n ; {\bf{p}}_1, \dots, {\bf{p}}_n\right)$,
indicated from now on by $ \left( s, m; {\bf{x}}, {\bf{p}} \right)$,
we define an operator $\hat{S}$ which constructs a new set of {\sl{three}} signed particles (in essence the same algorithm as in the single-body case
but different in practice because of the bigger phase-space involved)
$$
 \hat{S} \left[ \left( s, m; {\bf{x}}, {\bf{p}} \right)  \right] = \left\{ \left( s, m; {\bf{x}}', {\bf{p}}' \right), \left( +s, m; {\bf{x}}'', {\bf{p}}'' \right), \left( -s, m; {\bf{x}}''', {\bf{p}}''' \right) \right\}
$$
in the following manner:
\begin{itemize}
 \item
At time $t$, one generates a random number $r \in [0,1]$ and computes the
quantity $\delta t=-\frac{\ln{r}}{\gamma \left( {\bf{x}}(t)  \right)}$,
 \item
At time $t+\delta t$, the initial particle evolves as field-less, with new
coordinates $\left( {\bf{x}}', {\bf{p}}' \right) = \left( {\bf{x}}+\frac{{\bf{p}}(t)}{m} \delta t, {\bf{p}}  \right)$,
 \item
A pair of new signed particles is created at time $t+\delta t$, where the
particle with the sign $s$ has coordinates
$\left( {\bf{x}}'', {\bf{p}}'' \right) = \left( {\bf{x}}+\frac{{\bf{p}}(t)}{m} \delta t, {\bf{p}}+{\bf{p}}^*  \right) $ and the particle $-s$ has coordinates
$\left( {\bf{x}}''', {\bf{p}}''' \right) = \left( {\bf{x}}+\frac{{\bf{p}}(t)}{m} \delta t, {\bf{p}}-{\bf{p}}^*  \right) $
and the quantity ${\bf{p}}^*$ is computed from the normalized probability ${}^{\left| V_W \left({\bf{x}}; {\bf{p}} \right) \right|} /_{\gamma \left({\bf{x}}\right)}$.
\end{itemize}

Thus the operator $\hat{S}$  acts on a set of $n$ signed particles
(with $n$ an arbitrary natural number) by applying the above rules
to every single particle in the given ensemble. In a setting where
a time interval $\left[t,t+\Delta t \right]$ is implemented, the
operator $\hat{S}$ can be  applied iteratively to every particle
appearing in the creation process until the final time $t+\Delta t$
is reached. The final ensemble of particles consists of every particle
involved in the process described above (parents and created ones).
\footnote{The interested reader is encouraged to consult \cite{JCP-03}
and \cite{Shao} for details about the parallelization, complexity and
scaling of this algorithm.}

In this work we adhere to the work of Prof. D.~Ferry \cite{Ferry},
who considers a system as consisting of two Gaussian wave packets, denoted as
\begin{equation}
{\psi}_{T}(x)=\frac{1}{\sqrt{2}}\left [\psi_{1}(x-x_{0},p_{0})+\psi_{2}(x+x_{0},-p_{0})  \right ]
\label{function1}
\end{equation}
with one segment of the wave function being centered at $x_{0}$ moving in the positive direction,
while the other segment of the wave function  $-x_{0}$ moving in the negative direction. Accordingly,
the corresponding Wigner quasi-distribution function is represented by two traveling Gaussian wave
functions, with an "extra" term denoting the entanglement \cite{Ballentine}. Thus, 
the composite Wigner function can be described
as
\begin{equation}
f_{W}(x,p)\frac{1}{h}\left \{ e^{-(x-x_{0}^2/2\sigma^2-2\sigma^2(p-p_{0}^2)} + e^{-(x+x_{0})^2/2\sigma^2-2\sigma^2(p+p_{0}^2)} + 2e^{-x^2/2\sigma^2-2\sigma^2p^2} cos(x_{0}p) \right \} .
\label{function2}
\end{equation}
An interesting observation is the presence of an additional peak along the position and momentum
ones. The Gaussian peak, edging near $x=0$ and $p=0$, oscillates in the momentum position with
negation of the Gaussians. The existence of both positive and negative values in the
extra term indicates the lack of correspondence to a probability function. Interestingly, increasing
the $x_{0}$ term leads to rapid oscillation without any amplitude disturbance. 
As such, the Wigner function contributes to a richer context of the extra term,
which expresses a non-classical part of the distribution. 

Consequently, we consider a composite wave function such as (\ref{function1})
to describe an interaction between two systems for a specified period
of time, after which the interaction is eliminated. In other words, if a
system represented by two particles moving in opposite directions
after an interaction, with each system before and after the interaction
having different quasi-distribution functions, then (\ref{function1})
shall represent the description following the interaction of the
particles with a potential. 

Additionally, one could utilize the Bohm's interpretation of the
ERP system as a molecule of two atoms with a state of the total spin
equal to $0$ and the spin of each atom being $\frac{\hbar}{2}$ \cite{BohmERP}.
The model in (\ref{function2}) does not account for the spin of the
particles. On the other hand, the modern interpretation of the composite wave function 
combines the spin up and spin down wave functions and therefore producing the possibility
of the spin for each atom, essentially being an entangled wave function.
Moreover, once a measurement occurs, an information about a
single spin state is provided. Direct consequence of the entanglement of the wave
function is the inability to separate it into single wave functions for each particle.
Importantly, no measurement in the classical part can provide a form
for the entanglement, i.e. no quantum operant provides a real value
to accord for it. On the other hand, the extra term in (\ref{function2}) 
accounts for this provision. In fact, averaging the quasi-distribution function
over a quantity of some order accounts for the disappearance of the
extra term. It is entirely quantum mechanical term with no correspondence
to the classical physics.

We can furthermore explore the extra term of the function in terms of a
single Gaussian wave packet interacting with a tunneling barrier
\cite{shifren, kluksdahl, ravaioli} decaying into a pair of Gaussian packets
(in this case one is transmitted, and the other is reflected).
The corollary Wigner function exhibits the extra term, which is remaining
centered at the barrier. In a case where the term is negated and then time
is reversed, no convergence of particles into a single packet is observed. 
If the extra term is retained, a successful convergence
of the particles is present, signifying their correlation. 
Therefore, this event, preserving the information about the
creation of the particles fits the definition of entanglement \cite{shifren}.

Finally, several significant aspects should be commented on. First, the symbol
$\hat{S}$ is an actual operator in a strict mathematical sense. The
operator is defined over the space of tuples $\left( s, m; {\bf{x}}, {\bf{p}} \right)$,
mapped  over the space consisting of sets of three tuples (but one
can easily generalize this definition simply from the set of tuples
to the same set, by taking into account the fact that the operator
$\hat{S}$ is applied iteratively to every single signed particle).
Given a tuple, in this respect, its function consists of creating
a new set of three tuples. Secondly, it is feasible to show, in a
mathematical sense, that the operator is invertible. The reader
should recognize the fact that this operator is not necessarily
defined over an Hilbert space, although operators coming from the
standard (Schr\"{o}dinger) quantum mechanics always are.

Notably, while in the standard context, an operator can be interpreted
physically as the measurement of some macroscopic variable, in the
signed particle formulation, the operator $\hat{S}$ is not concerned
with any measurement at all. Rather, it represents a completely
different mathematical object driving the evolution of a system.

While this operator is not necessarily Hermitian (in a strictly
mathematical sense, i.e. $\hat{S}* \hat{S}$ is not necessarily
equal to the identity operator $\hat{I}$), it is certainly a
unitary operator in a physical sense. Broadly speaking, in a
physical sense a unitary operator in quantum mechanics could
be simply seen as an mathematical object which assures that,
in a $N$-body system, all probabilities in the configuration
space must sum up to exactly the value $N$. In this sense,
the operator $\hat{S}$ (or, more specifically, the iterative
application of $\hat{S}$) is a unitary operator.

\section*{\sffamily \Large Entanglement and signed particles}

Up to date, from a purely physics standpoint, entanglement remains a very puzzling concept which has no classical
analogue, therefore producing counterintuitive conclusions \cite{Einstein}.
The concept of entanglement can be perceived (in a broad sense) as the incapacity of a quantum many-body
system to be described as an ensemble of perfectly independent bodies (and, therefore, only a truly many-body
approach can handle such problems). From a mathematical standpoint, one may see as an equivalent illustration 
the  impossibility of expressing the quantum state of the entire system in terms of a product of
single-body states, or more formally, to say that the phase-space surface or manifold generated by
the quasi-distribution function of any entangled system cannot be decomposed into a product of lower
dimensional surfaces \cite{Almeida}.

\begin{figure}[h!]
\centering
\begin{minipage}{1.0\textwidth}
\centering
\begin{tabular}{c}
\includegraphics[width=0.5\textwidth]{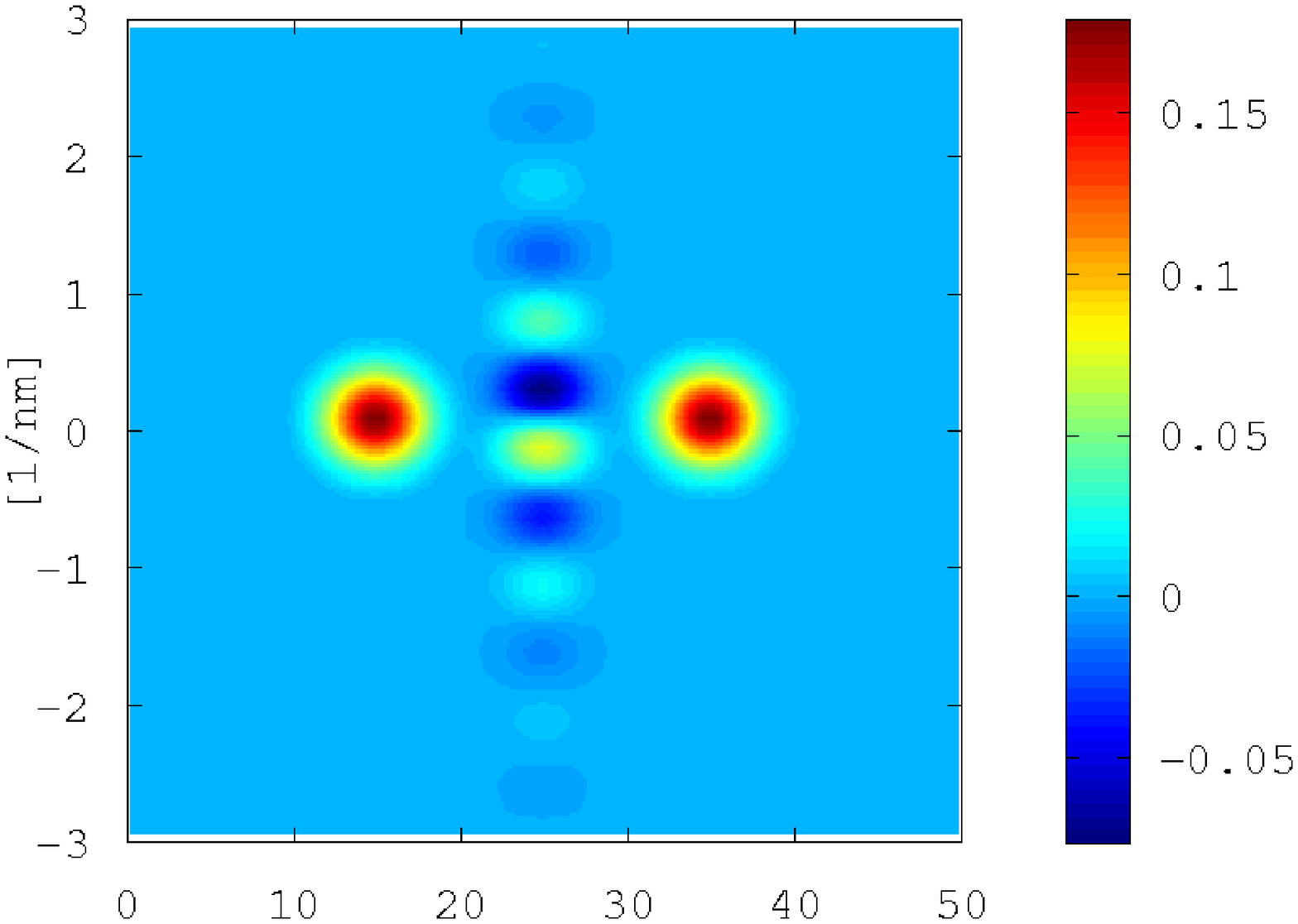}
\includegraphics[width=0.5\textwidth]{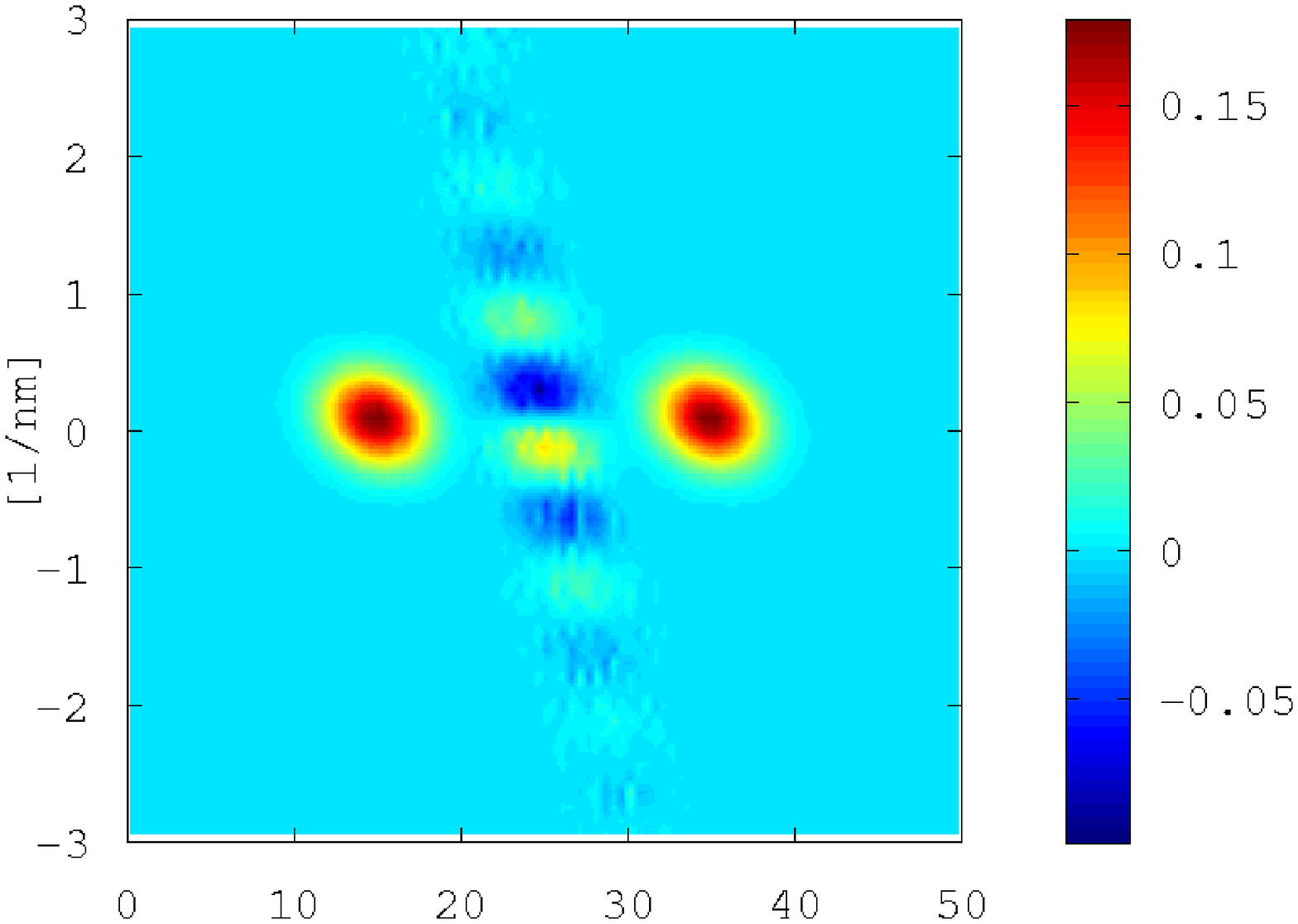}
\\
\includegraphics[width=0.5\textwidth]{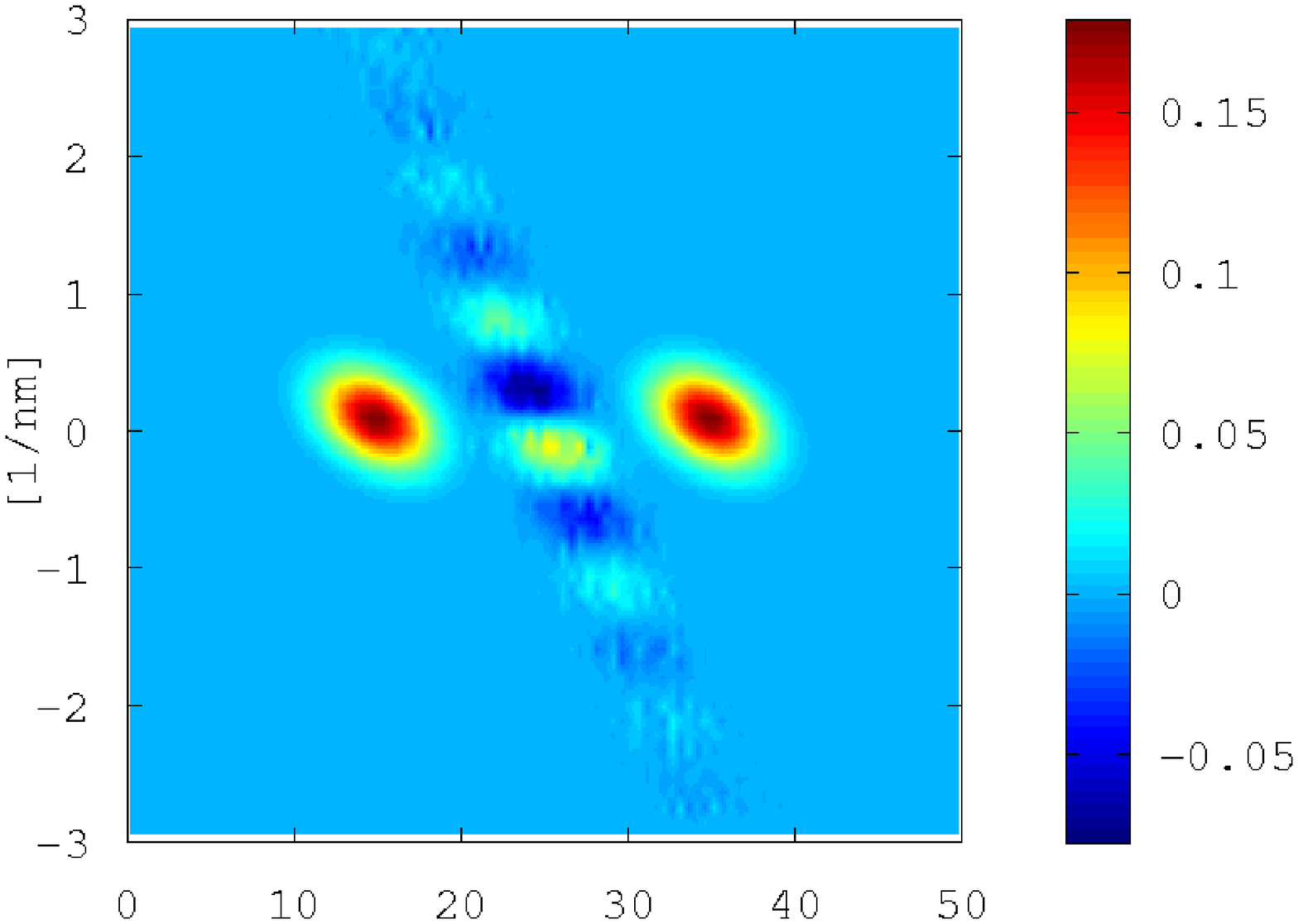}
\includegraphics[width=0.5\textwidth]{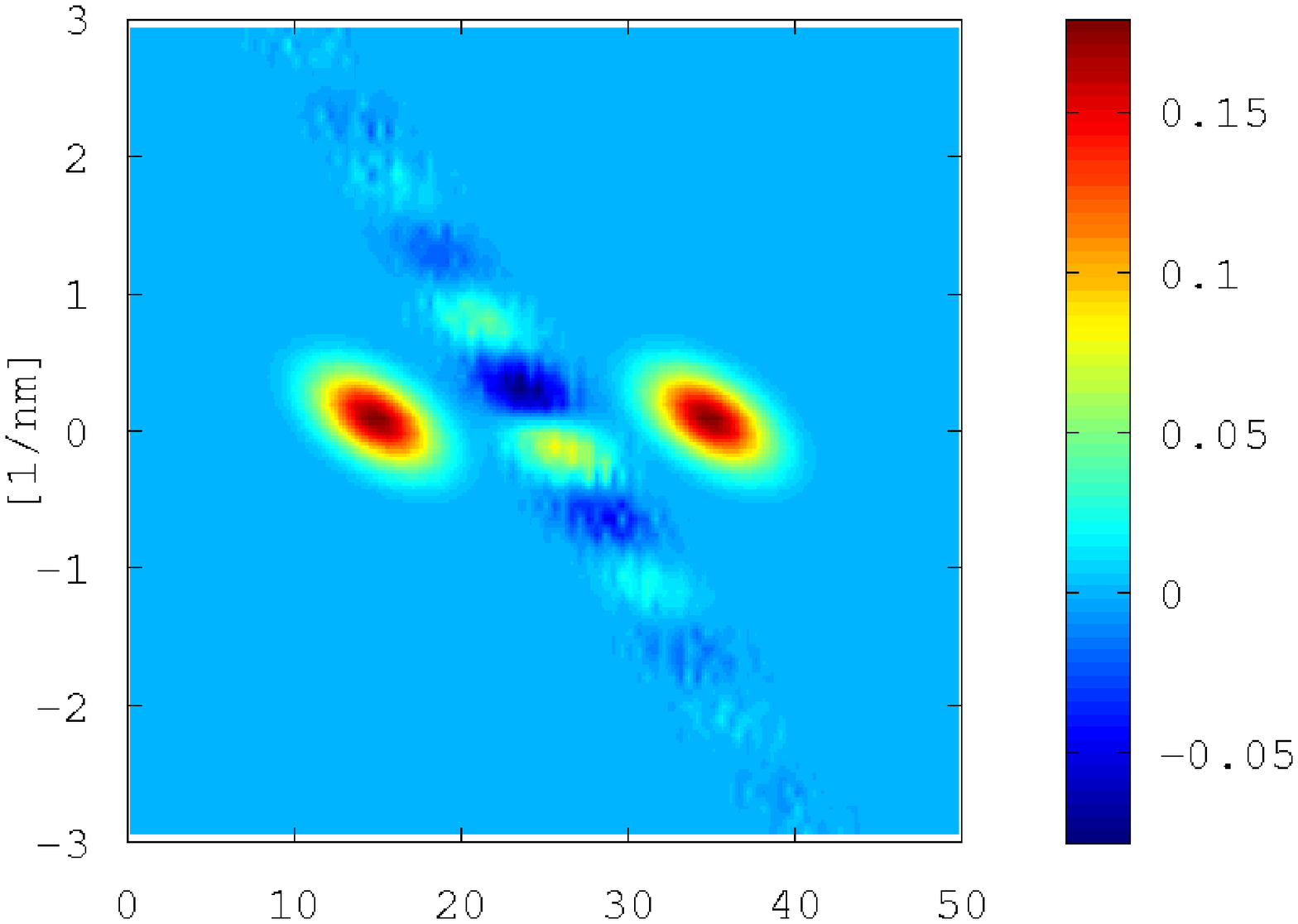}
\end{tabular}
\end{minipage}
\caption{Evolution of an entangled two-body system in the ballistic case. The (normalized) quasi-distribution function in the (reduced) phase-space are shown at times $0$ fs, $1$ fs, $2$ fs and $3$ fs respectively (from top to bottom, left to right). The entanglement oscillations remains stable during the simulation and only rotate (as expected).}
\label{ballistic_2.5_1.5}
\end{figure}

In spite of the (relative) simplicity of the available mathematical definitions of quantum entanglement,
the physical world of quantum entanglement still exhibits puzzling peculiarities. One of the most difficult
and at the same time fundamental endeavors is how to {\sl{intuitively understand}} entanglement,
particularly in the case when a non-phase-space approach is adopted. Indeed, it remains challenging
to discern whether two objects are entangled only from the mathematical expression of a corresponding
two-body wave-function. Interestingly, the adoption of a phase-space approach to quantum mechanics
can provide a rather perceptive understanding of the occurrence in that the entanglement between
two particles can be intuitively thought as rapid (negative and positive) oscillations in the
phase-space (e.g. see Fig. \ref{ballistic_2.5_1.5}). Therefore, despite it simplicity, this
is the (intuitive) picture considered (and exploited) in this section.

\bigskip

Consequently, in a four-dimensional phase-space, the initial conditions for a two-body system consisting of two strongly
entangled bodies (e.g. electrons) could be expressed as (which corresponds to a physically realizable system
as they fulfill the Tatarskii conditions \cite{tatarskii}):
\begin{eqnarray}
 f_W^0 \left(x_1, x_2; p_1, p_2 \right) &=& C e^{-\left( \frac{x_1-x_1^0}{\sigma_1^x} \right)^2} e^{-\left( \frac{p_1-p_1^0}{\sigma_1^p} \right)^2}
 e^{-\left( \frac{x_2-x_2^0}{\sigma_2^x} \right)^2} e^{-\left(\frac{p_2-p_2^0}{\sigma_2^p} \right)^2} \nonumber \\
&+& C e^{-\left( \frac{x_1-x_1^0}{\sigma^x_{ent}} \right)^2 } e^{-\left( \frac{x_2-x_2^0}{\sigma^x_{ent}} \right)^2 } e^{-\frac{ \left| {\bf{p}}-{\bf{p}}_0 \right| }{\sigma_0}}  \nonumber \\
 & \times &2 \sin{\left( \frac{p_1-p_1^0}{\sigma^p_{ent}} \right)} \nonumber \\
 & \times &2 \sin{\left( \frac{p_2-p_2^0}{\sigma^p_{ent}} \right)},
\label{entangled}
\end{eqnarray}
where $C$ is a normalization constant and $x_1^0$, $x_2^0$, $p_1^0$, $p_2^0$, etc, are parameters defining the exact shape of the entanglement
from a phase-space perspective. Importantly the extra term here depicts a system of entangled particles, in 
accordance with the explanation provided by prof. D.~Ferry. Mathematically speaking, one could see the initial
conditions (\ref{entangled}) as attained from a generalization of the quasi-distribution obtained from the
application of the Wigner-Weyl transform to the two-body wave-function coming from the superposition of two
Gaussian single-body wave-functions (and the reader should note that, although a common practice, it is not
always necessary to start from the concept of a wave-function).

In the particular numerical experiments performed in this work - 
$x_1^0=15$ nm, $x_2^0=35$ nm (initial positions of the peak of the two Gaussian wave-packets in the configuration space),
$p_1^0=p_2^0=0$ nm$^{-1}$ (initial positions of the peak of the two Gaussian wave-packets in the momentum space),
$\sigma_1^x = \sigma_2^x = 3$ nm (spatial dispersion of the two Gaussian wave-packets),
$\sigma_1^p = \sigma_2^p = 1/3$ nm$^{-1}$ (momentum dispersion of the two Gaussian wave-packets),
$\sigma_0 = 0.75$ nm$^{-1}$, the parameters $\sigma^x_{ent}$ and $\sigma^p_{ent}$ are fixed case by case (see below).
How these oscillations can be affected by an external dissipative environment is the topic of the next subsection.
Therefore, in the following, we describe several numerical simulations, performed by applying the many-body
signed particle formulation of quantum mechanics, which aim is to understand what can eventually strengthen or weaken
the entanglement between particles.

\bigskip

{\sffamily \Large Numerical experiments}

The current section examines a system of two distinguishable and non-interacting\footnote{Manifestedly, the
more naturalistic condition would involve the simulation of indistinguishable electrons interacting
with each other by means of their long-range Coulomb potential. For the sake of simplicity, in the
current study, we ignore these interactions and do not take into account the indistinguishability as well.}
entangled electrons, immersed in a dissipative background. The effect on a signed particle is reduction
of its energy by a random (but controlled) amount. In this occurrence, for a signed particle with a
(one-dimensional) momentum $p$, the following algorithm is applied: assuming two fixed (and user specified)
numbers $r_{prob} \in [0, 1]$ and $r_{\%} \in ]0,1]$, we generate a random number $r_1$, homogeneously
distributed in the range $[0, 1]$, and if $r_1<r_{prob}$ then the momentum of the particle is set to
$p^{new} = (1 - r_{\%}) \times p$ (therefore reducing the energy of the particle). 
To be more precise, there are two principal reasons to focus on this oversimplified situation.
Notwithstanding the existence of more realistic models for dispersive baths
for the Wigner formalism, they are all defined for the single-body case, and to the best of
our knowledge many-body scattering terms have not been formulated yet.
Additionally, the main goal of this paper is to focus on a model
which main features are expected to be valid also for eventual many-body models. \footnote{The
reader should note that this should be considered as a first approach to the problem
of many-body systems immersed in a dispersive background (or bath). Indeed, the
conclusions drawn in this study will be invalid in the context of a more realistic model.}
Imperatively, despite its simplicity, this model is able to introduce a controlled amount
of dissipation (or noise) in the system via a strategy, hardly achievable if other
formulations of quantum mechanics are considered. 

\begin{figure}[h!]
\centering
\begin{minipage}{1.0\textwidth}
\centering
\begin{tabular}{c}
\includegraphics[width=0.5\textwidth]{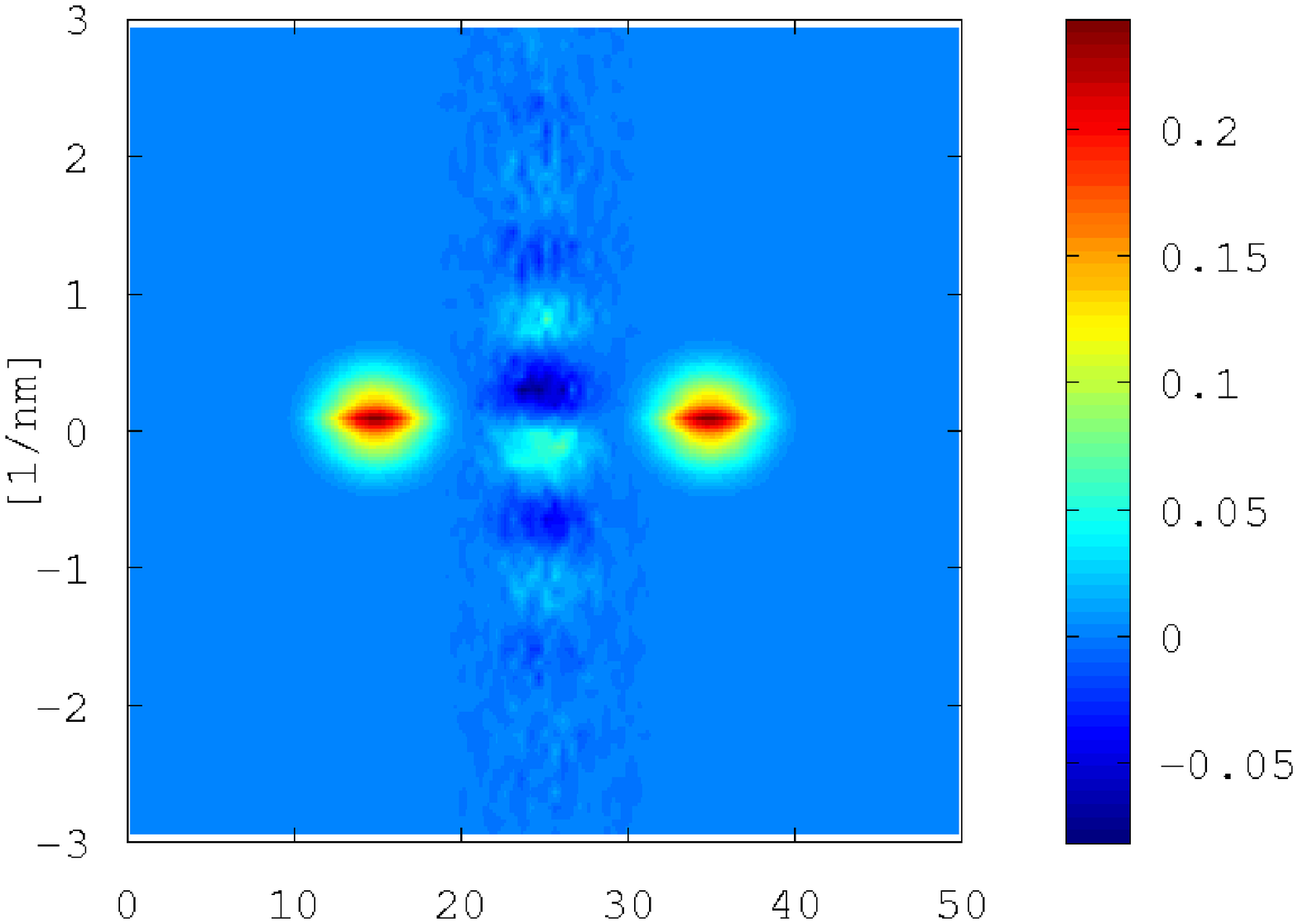}
\includegraphics[width=0.5\textwidth]{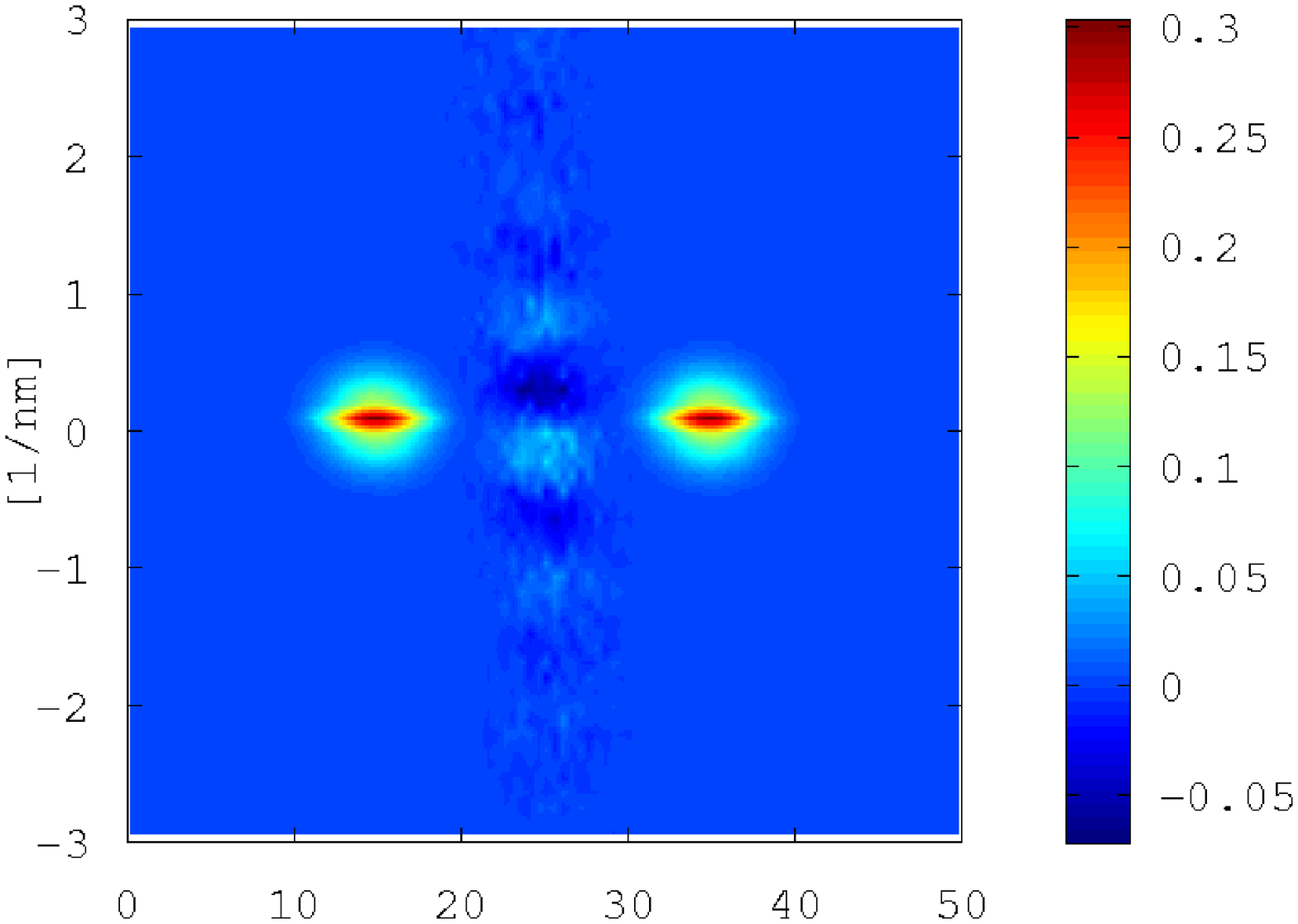}
\\
\includegraphics[width=0.5\textwidth]{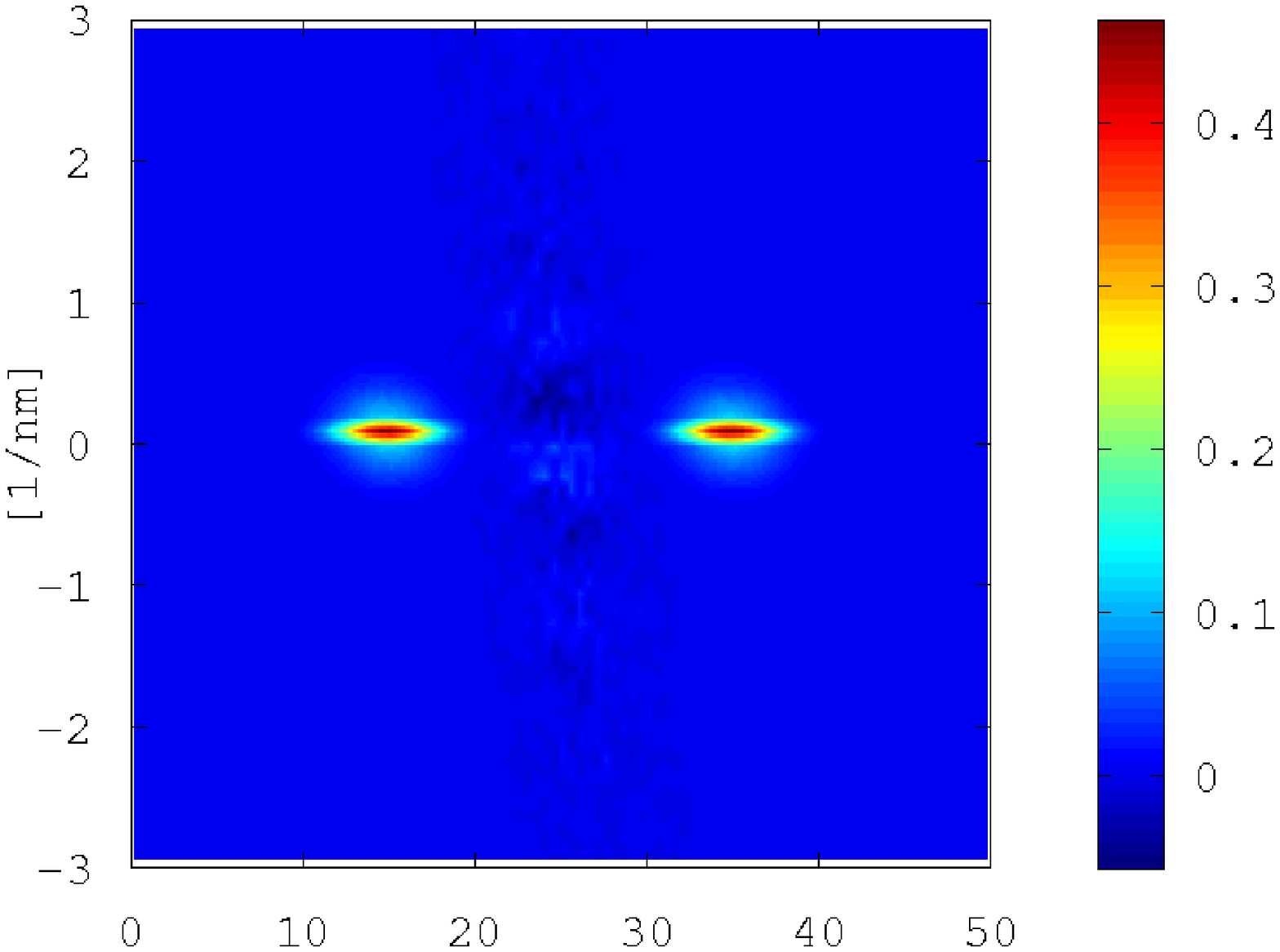}
\includegraphics[width=0.5\textwidth]{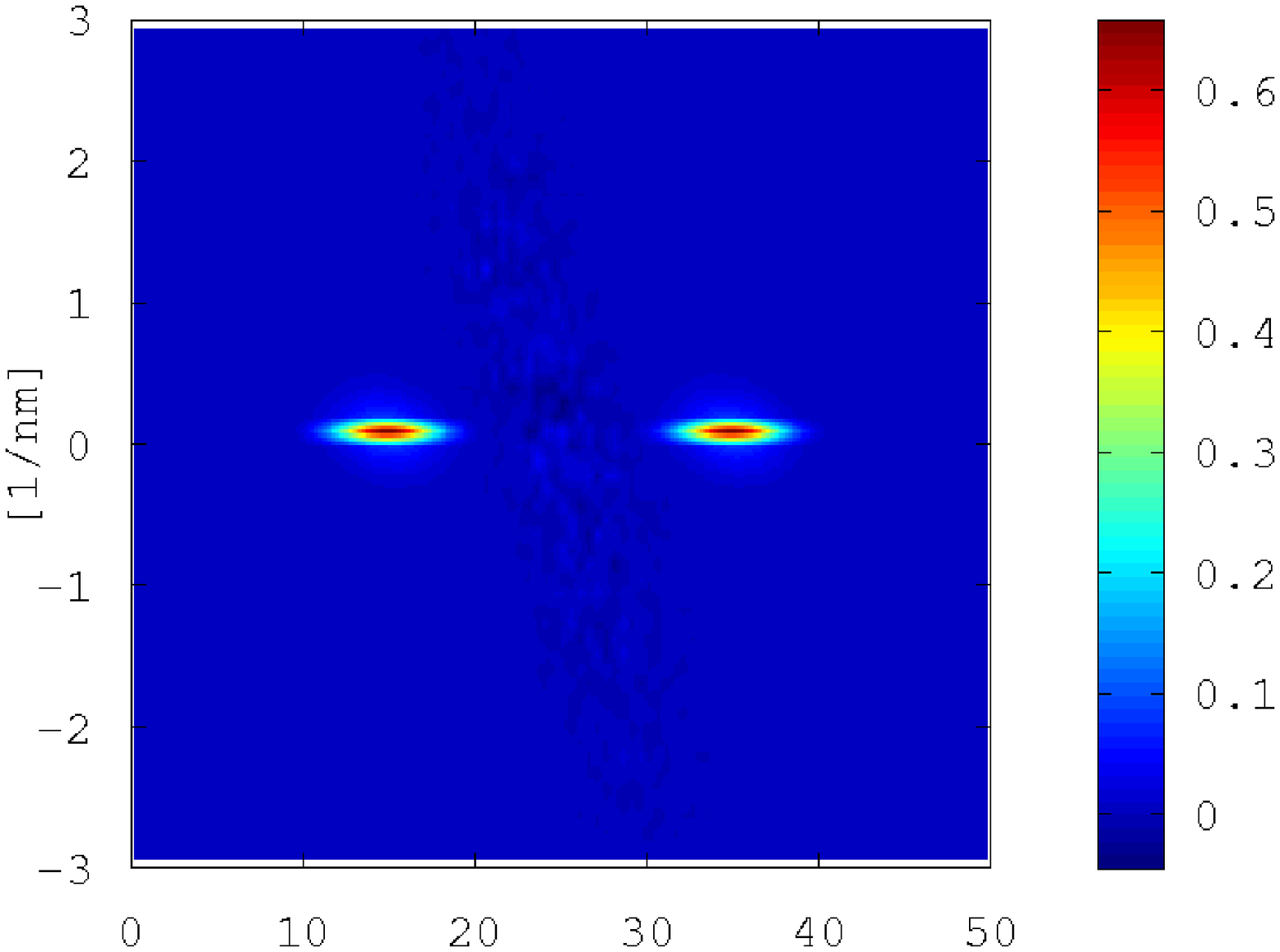}
\end{tabular}
\end{minipage}
\caption{Evolution of an entangled two-body system in the presence of a dissipative background ($2$\% probability of scattering, removing $15$\% of the energy from a particle, $\sigma_{ent}^x = 2.5$ nm, $\sigma^p_{ent} = \frac{3}{2} \Delta p$ with $\Delta p = \frac{\hbar \pi}{30.0}$ nm$^{-1}$). The (normalized) quasi-distribution function in the (reduced) phase-space are shown at times $0.1$ fs, $0.2$ fs, $0.5$ fs and $1$ fs respectively (from top to bottom, left to right). Despite the small amount of dissipation, the entanglement oscillations are destroyed quickly and completely disappear at $1$ fs.}
\label{distribution_0.02_0.85_2.5_1.5}
\end{figure}

\begin{figure}[h!]
\centering
\begin{minipage}{1.0\textwidth}
\centering
\begin{tabular}{c}
\includegraphics[width=0.5\textwidth]{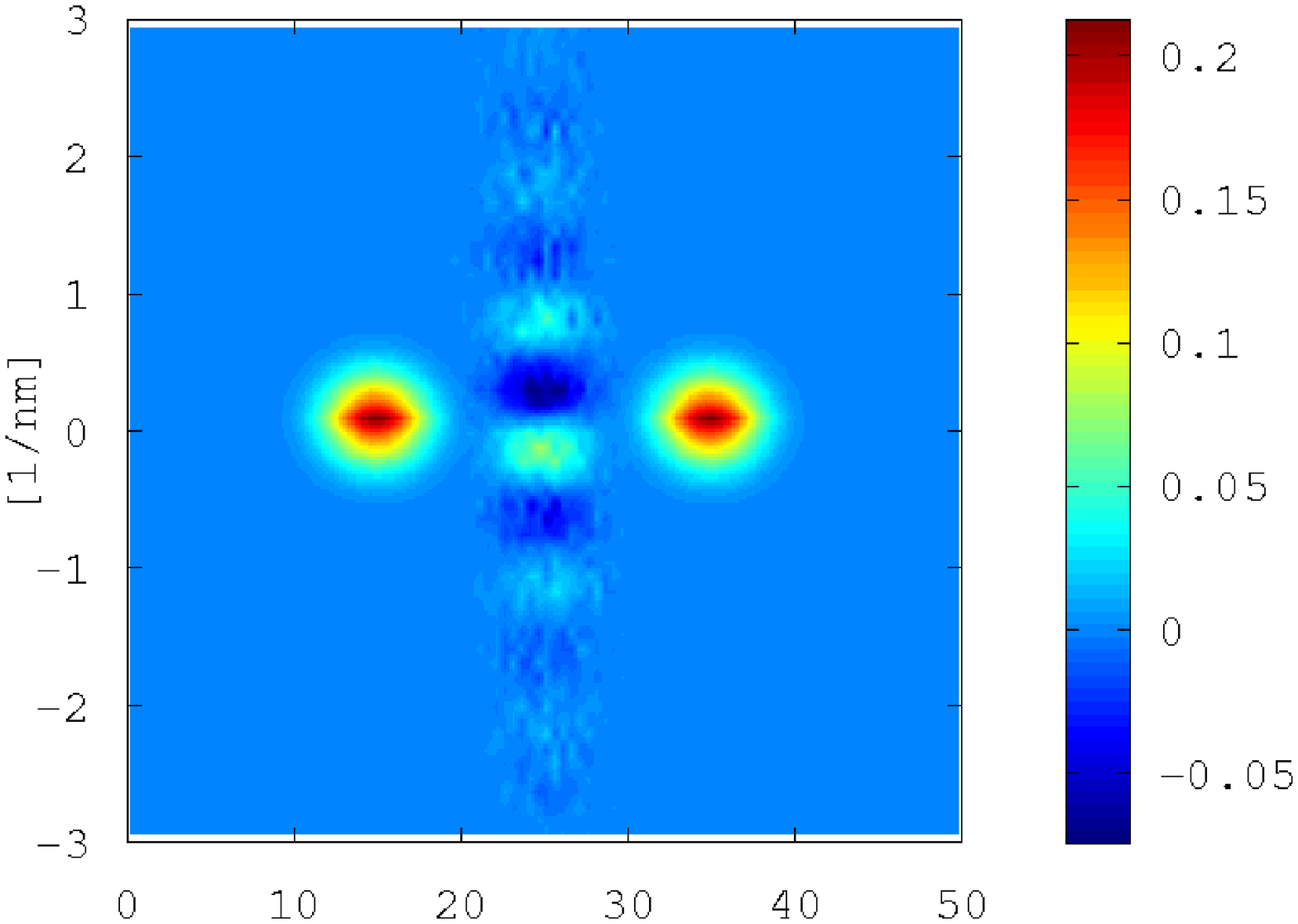}
\includegraphics[width=0.5\textwidth]{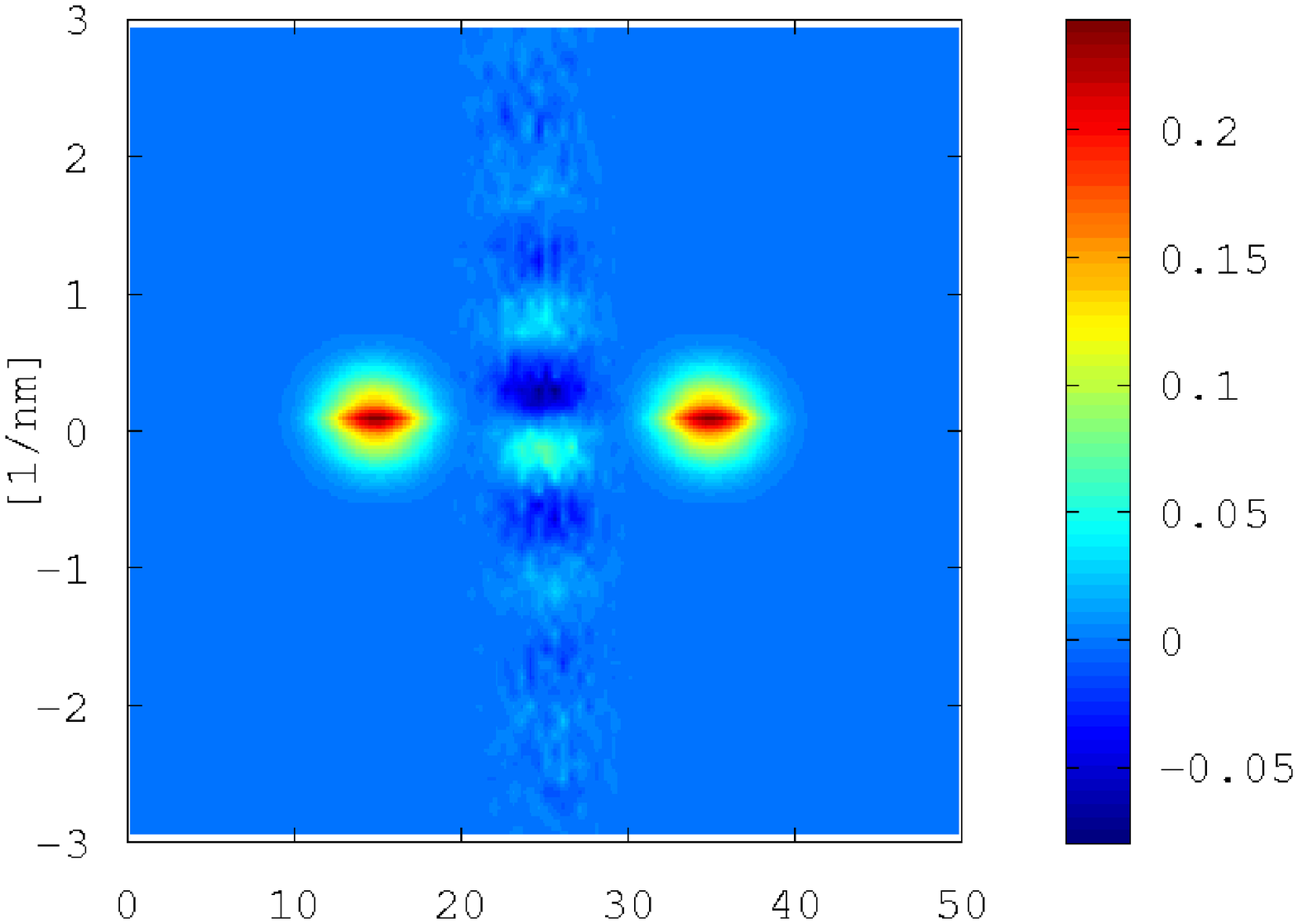}
\\
\includegraphics[width=0.5\textwidth]{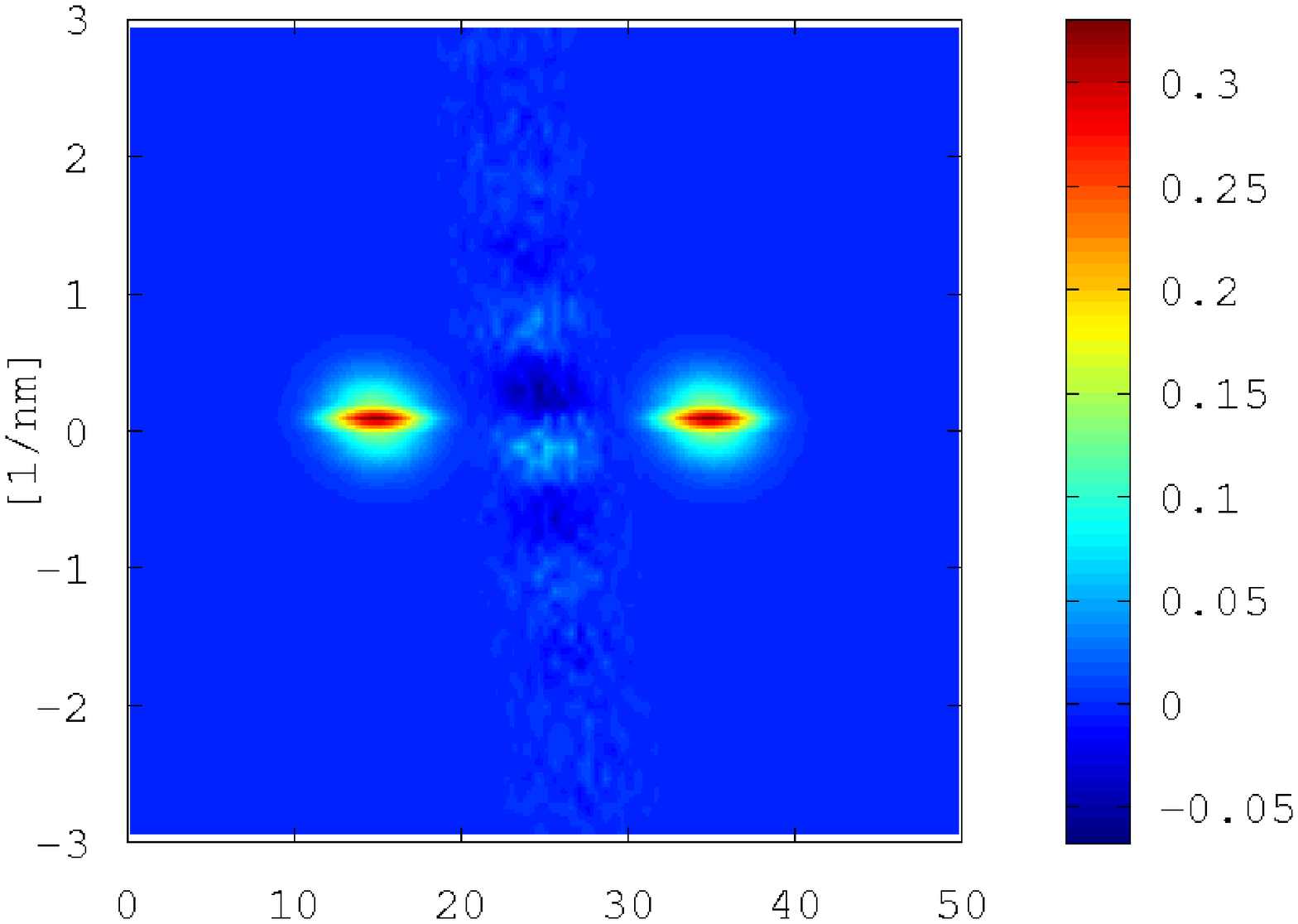}
\includegraphics[width=0.5\textwidth]{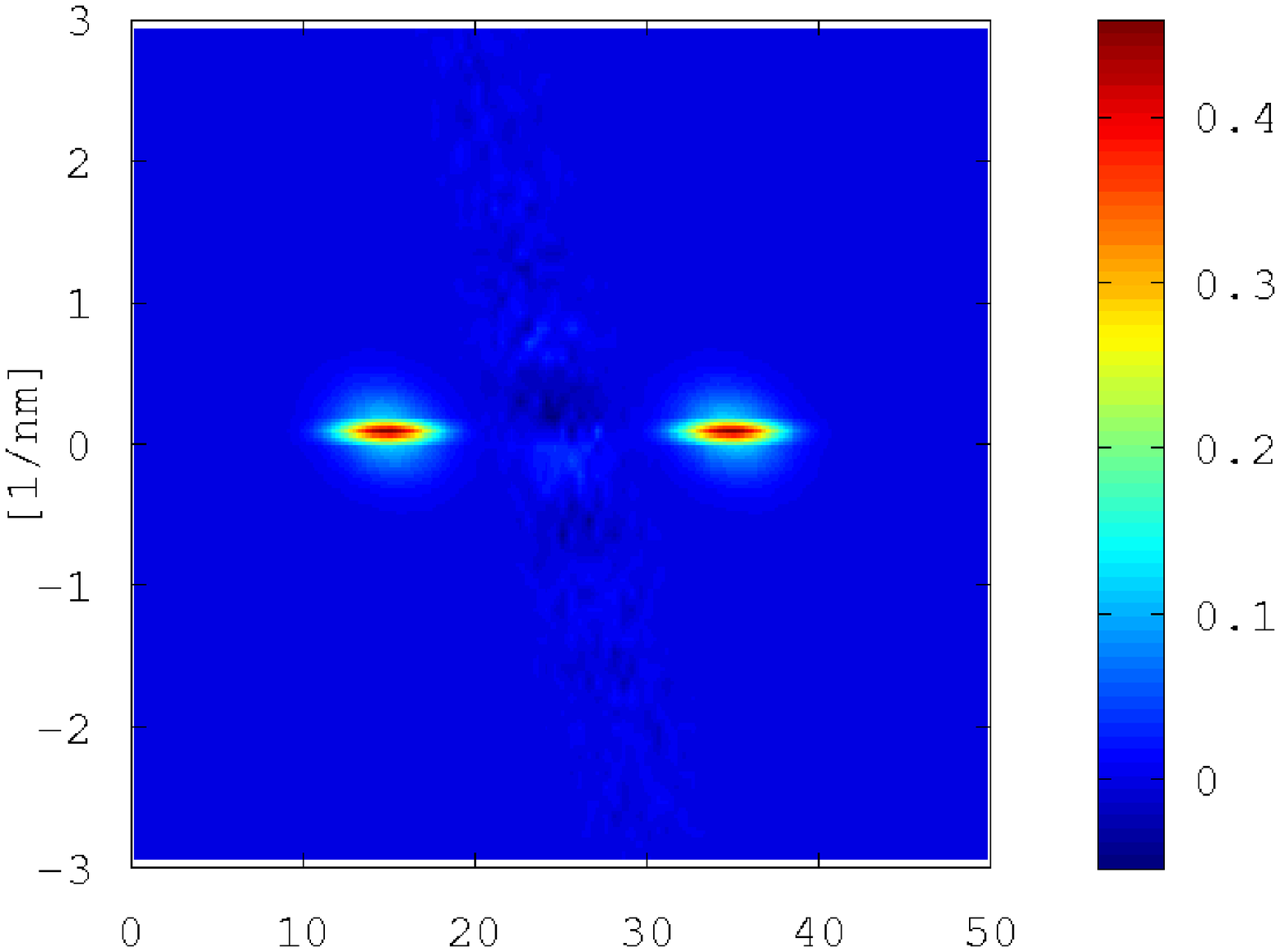}
\end{tabular}
\end{minipage}
\caption{Evolution of an entangled two-body system in the presence of a dissipative background ($1$\% probability of scattering, removing only $5$\% of the energy from a particle, $\sigma_{ent}^x = 2.5$ nm, $\sigma^p_{ent} = \frac{3}{2} \Delta p$ with $\Delta p = \frac{\hbar \pi}{30.0}$ nm$^{-1}$). The (normalized) quasi-distribution function in the (reduced) phase-space are shown at times $0.1$ fs, $0.2$ fs, $0.5$ fs and $1$ fs respectively (from top to bottom, left to right). Being the amount of dissipation smaller than in Fig. \ref{distribution_0.02_0.85_2.5_1.5}, the entanglement oscillations still decays with time but at a slower rate. In fact, at time $0.5$ fs the oscillations due to entanglement are still clearly observable.}
\label{distribution_0.01_0.95_2.5_1.5}
\end{figure}

\begin{figure}[h!]
\centering
\begin{minipage}{1.0\textwidth}
\centering
\begin{tabular}{c}
\includegraphics[width=0.45\textwidth]{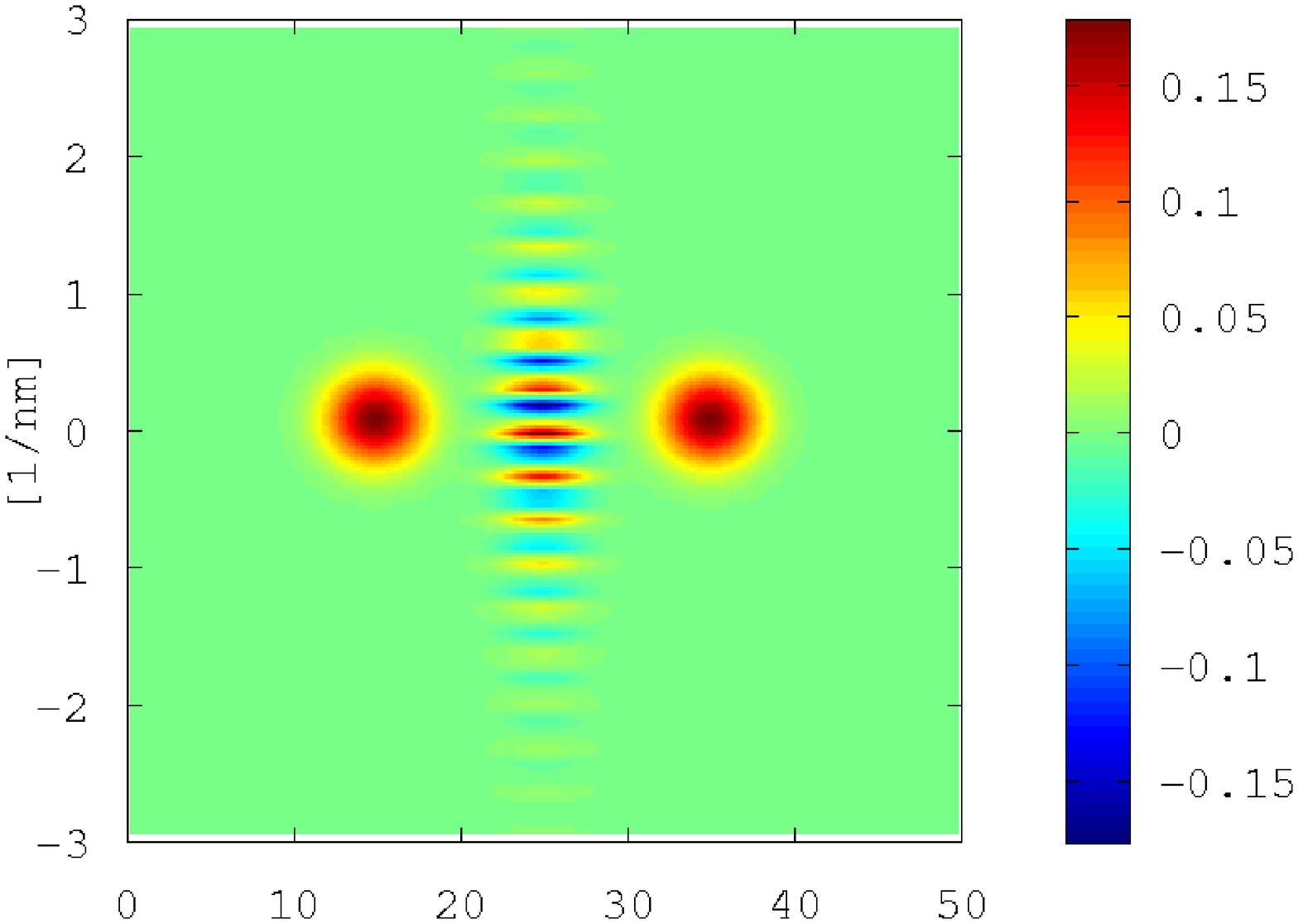}
\\
\includegraphics[width=0.45\textwidth]{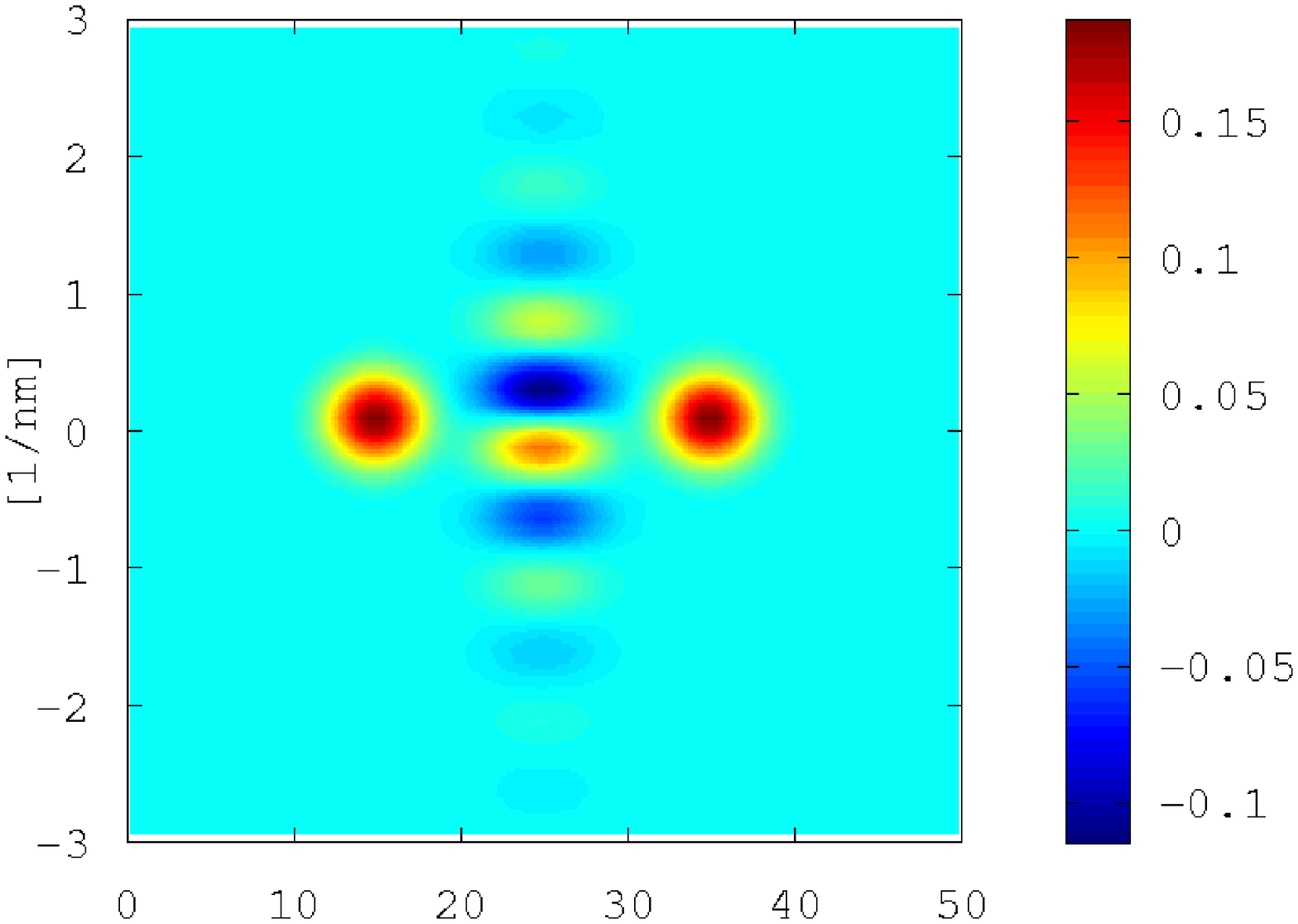}
\\
\includegraphics[width=0.45\textwidth]{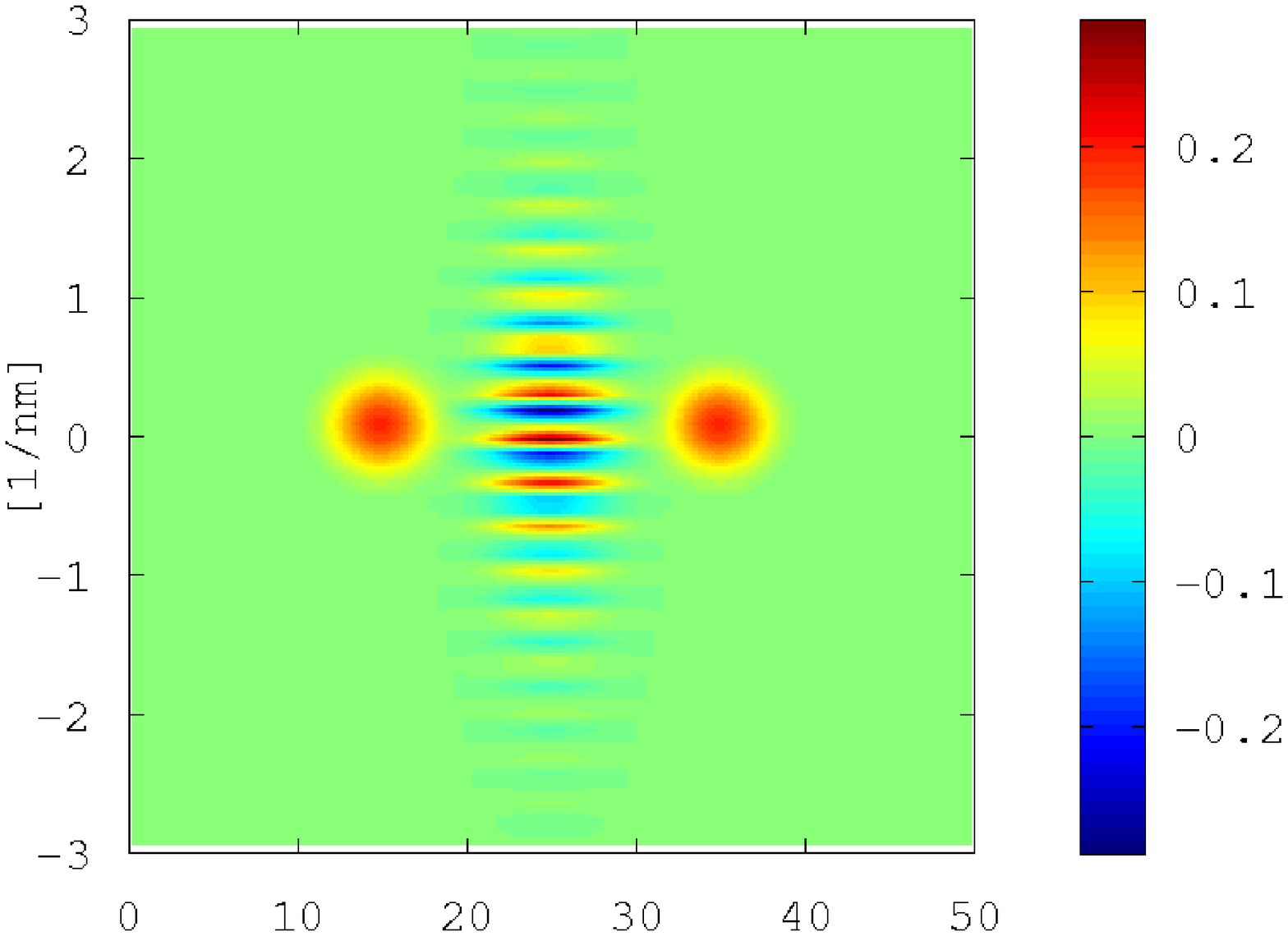}
\end{tabular}
\end{minipage}
\caption{Initial conditions of entangled two-body systems with (top) $1$\% probability of scattering, removing only $5$\% of the energy from a particle, $\sigma_{ent}^x = 2.5$ nm, and $\sigma^p_{ent} = \frac{\Delta p}{2}$, with (middle) $1$\% probability of scattering, removing only $5$\% of the energy from a particle, $\sigma_{ent}^x = 3.5$ nm, and $\sigma^p_{ent} = \frac{3}{2} \Delta p$, and with (bottom) $1$\% probability of scattering, removing only $5$\% of the energy from a particle, $\sigma_{ent}^x = 3.5$ nm, and $\sigma^k_{ent} = \frac{\Delta p}{2}$ respectively.}
\label{initial_distributions}
\end{figure}

\begin{figure}[h!]
\centering
\begin{minipage}{1.0\textwidth}
\centering
\begin{tabular}{c}
\includegraphics[width=0.5\textwidth]{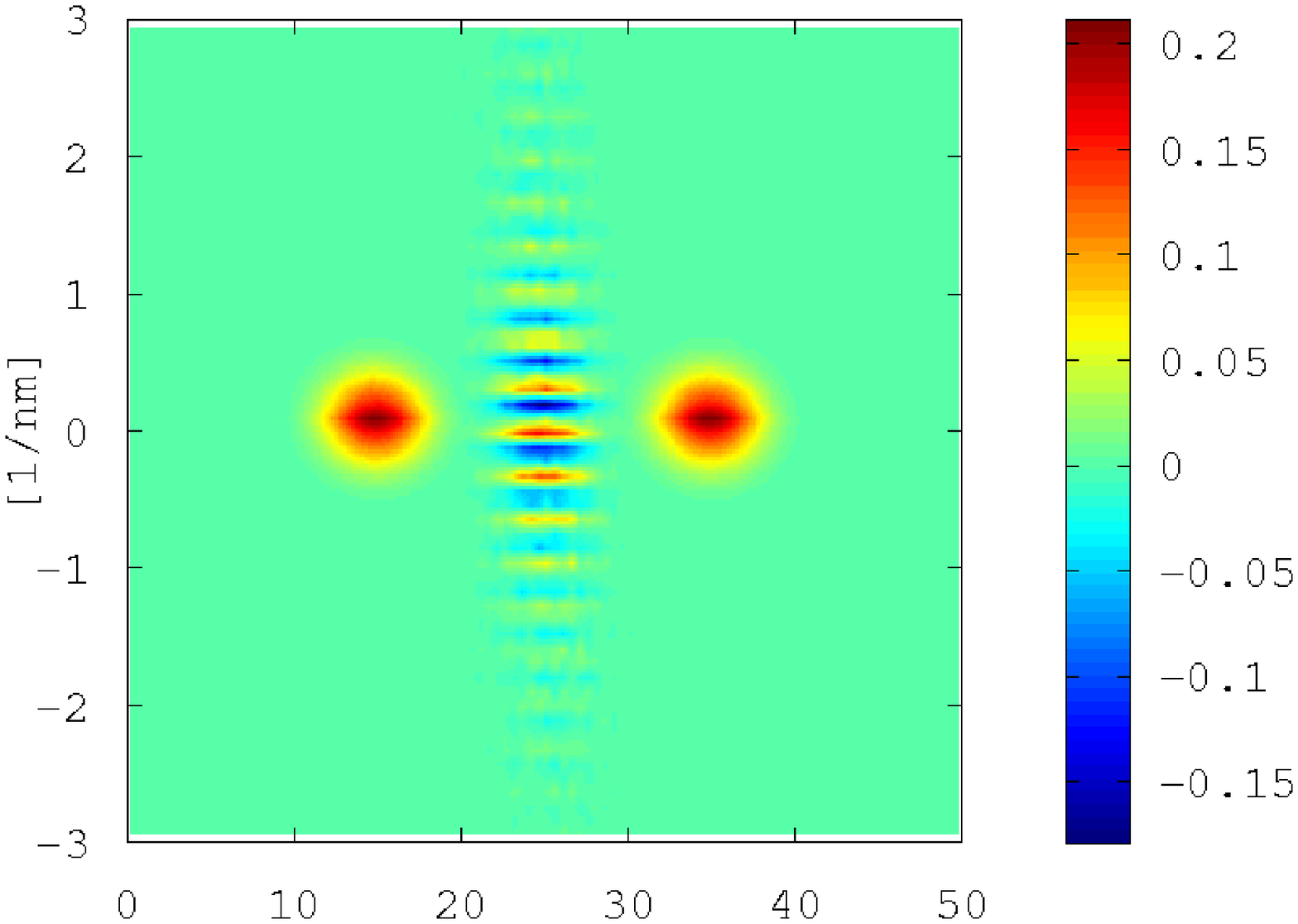}
\includegraphics[width=0.5\textwidth]{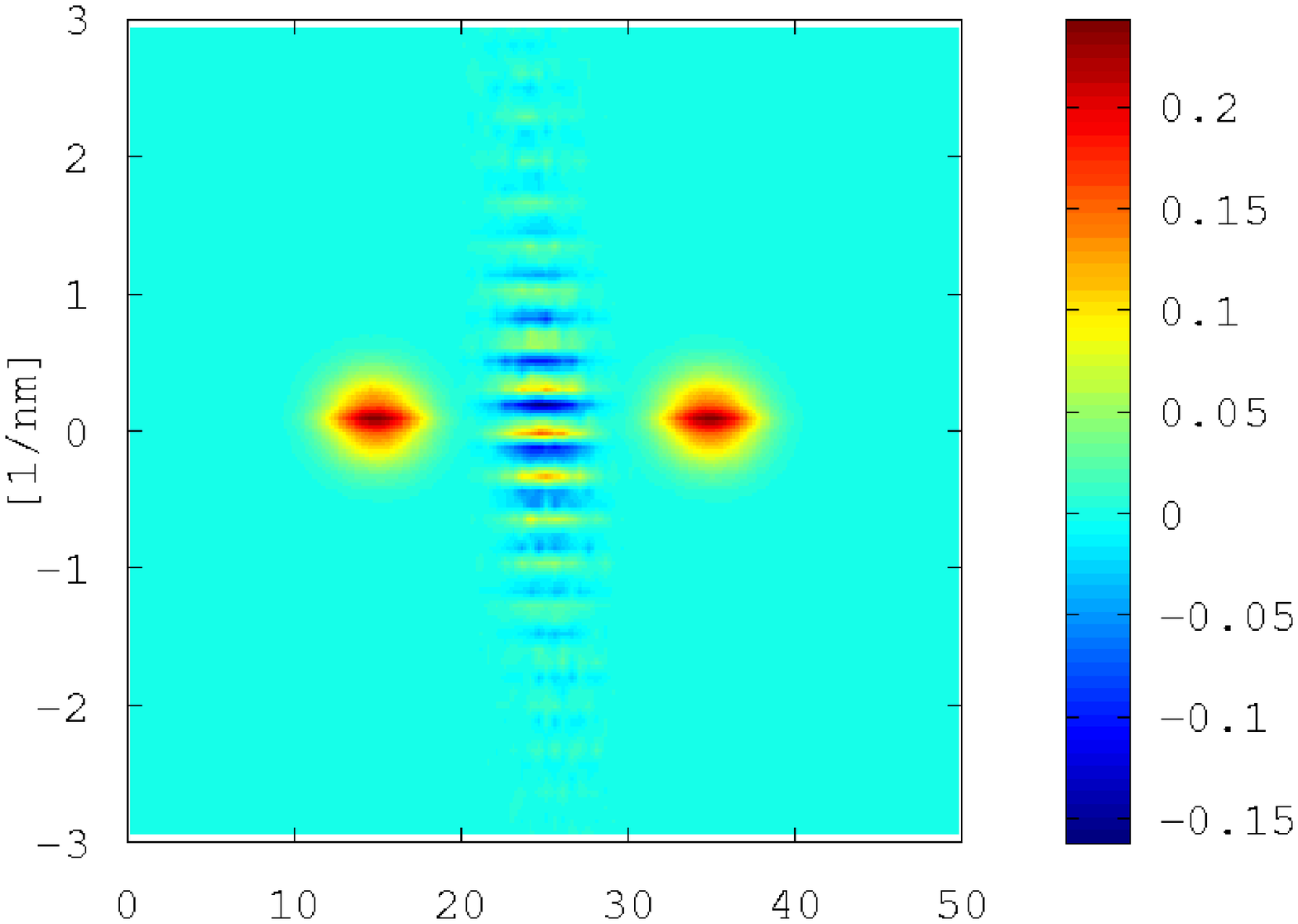}
\\
\includegraphics[width=0.5\textwidth]{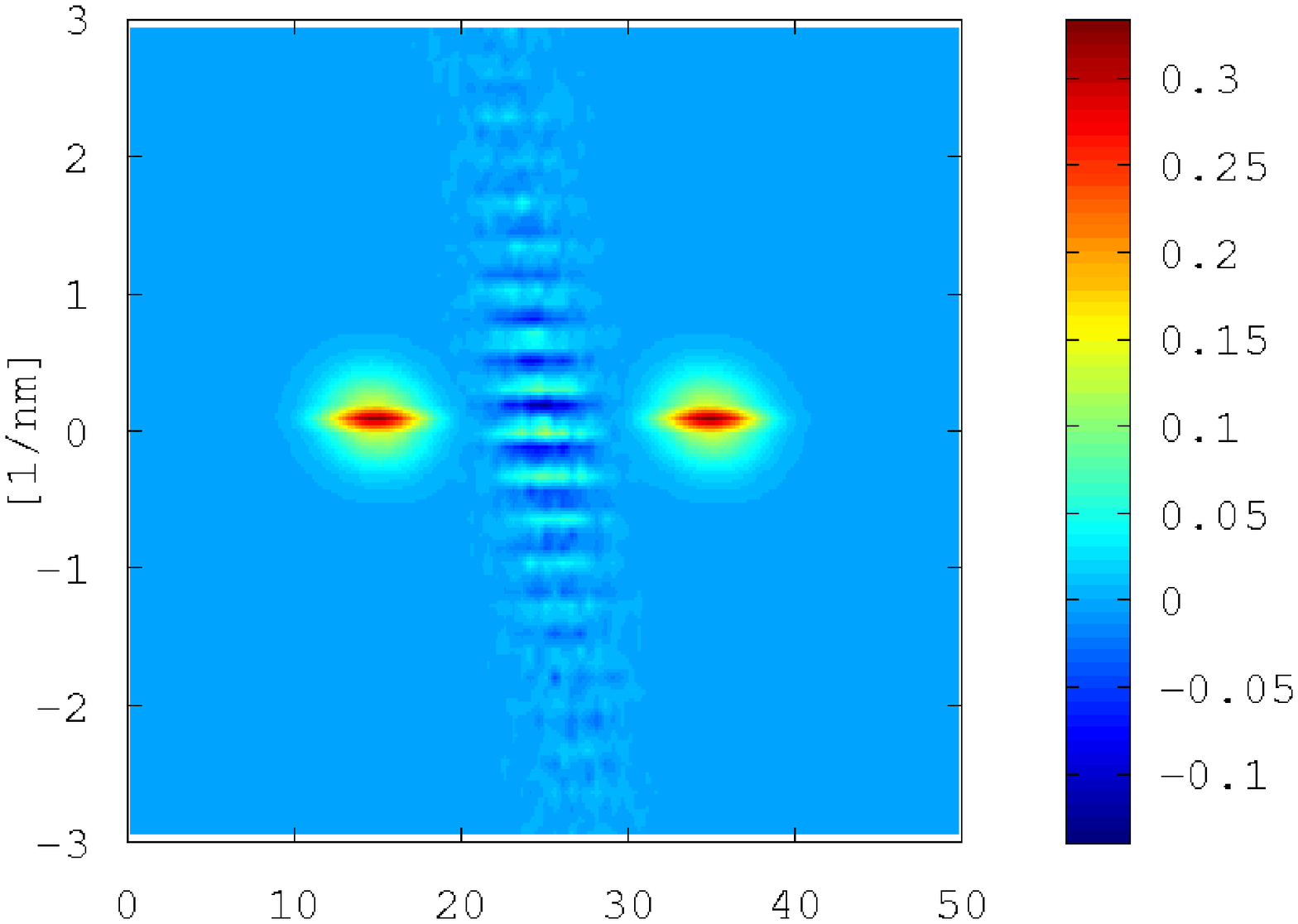}
\includegraphics[width=0.5\textwidth]{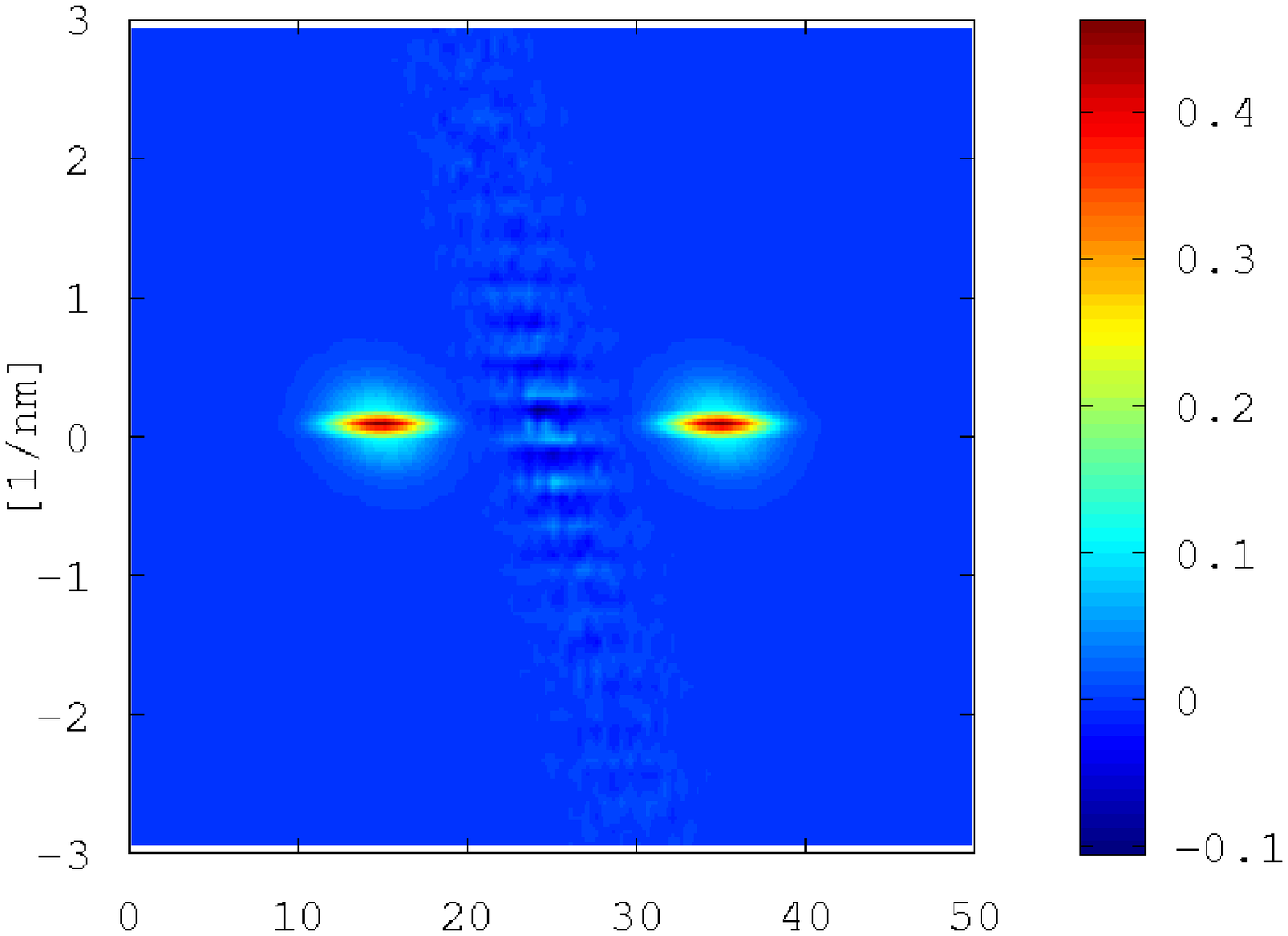}
\end{tabular}
\end{minipage}
\caption{Evolution of an entangled two-body system in the presence of a dissipative background ($1$\% probability of scattering, removing only $5$\% of the energy from a particle, $\sigma_{ent}^x = 2.5$ nm, $\sigma^p_{ent} = \frac{\Delta p}{2}$). The (normalized) quasi-distribution function in the (reduced) phase-space are shown at times $0.1$ fs, $0.2$ fs, $0.5$ fs and $1$ fs respectively (from top to bottom, left to right). While the amount of dissipation is the same as in Fig. \ref{distribution_0.01_0.95_2.5_1.5}, the entanglement oscillations decays slowerly than in Fig. \ref{distribution_0.01_0.95_2.5_1.5}, due to the different initial oscillation pattern in the entanglement (see top of Fig. \ref{initial_distributions}).}
\label{distribution_0.01_0.95_2.5_0.5}
\end{figure}

\begin{figure}[h!]
\centering
\begin{minipage}{1.0\textwidth}
\centering
\begin{tabular}{c}
\includegraphics[width=0.5\textwidth]{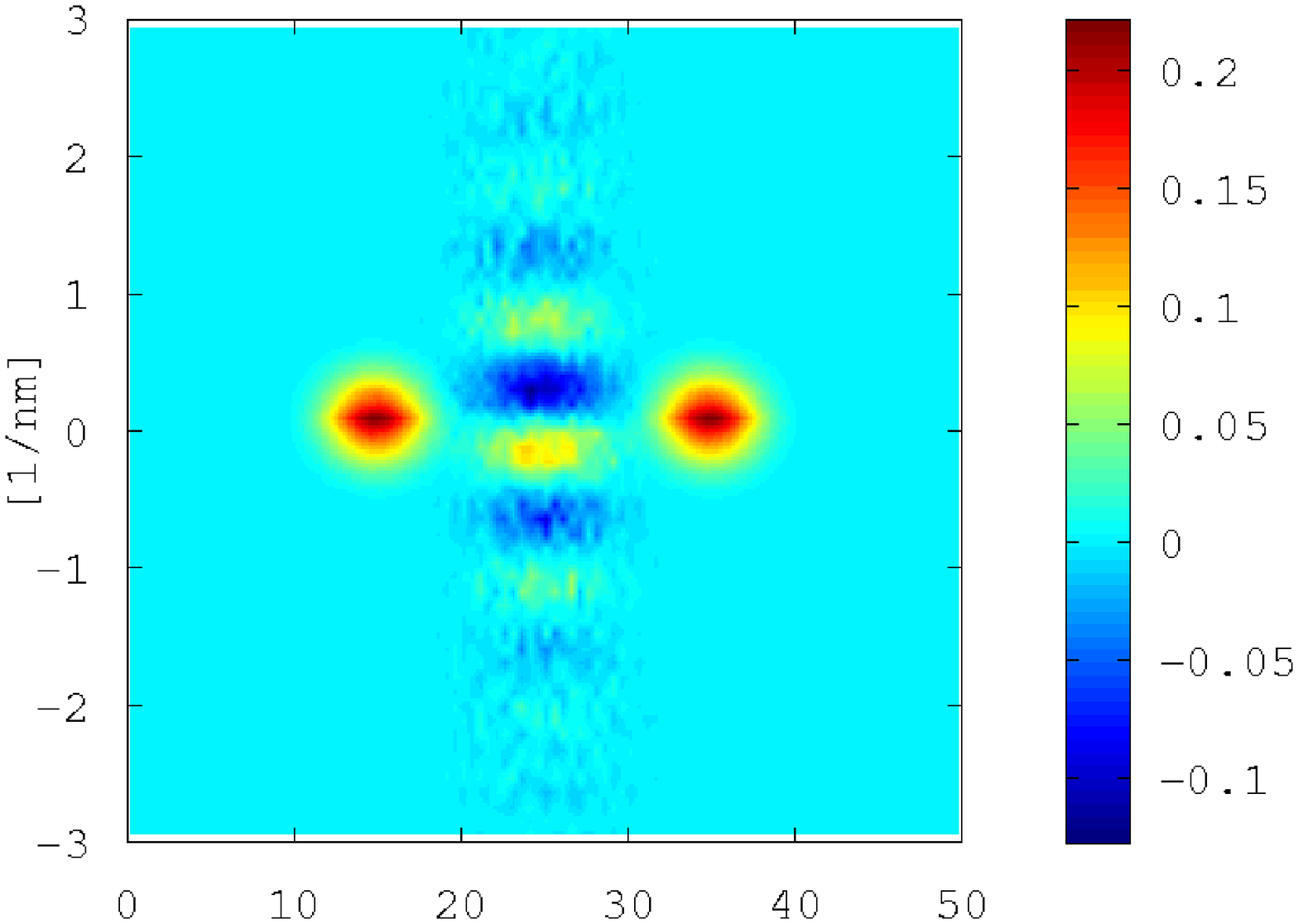}
\includegraphics[width=0.5\textwidth]{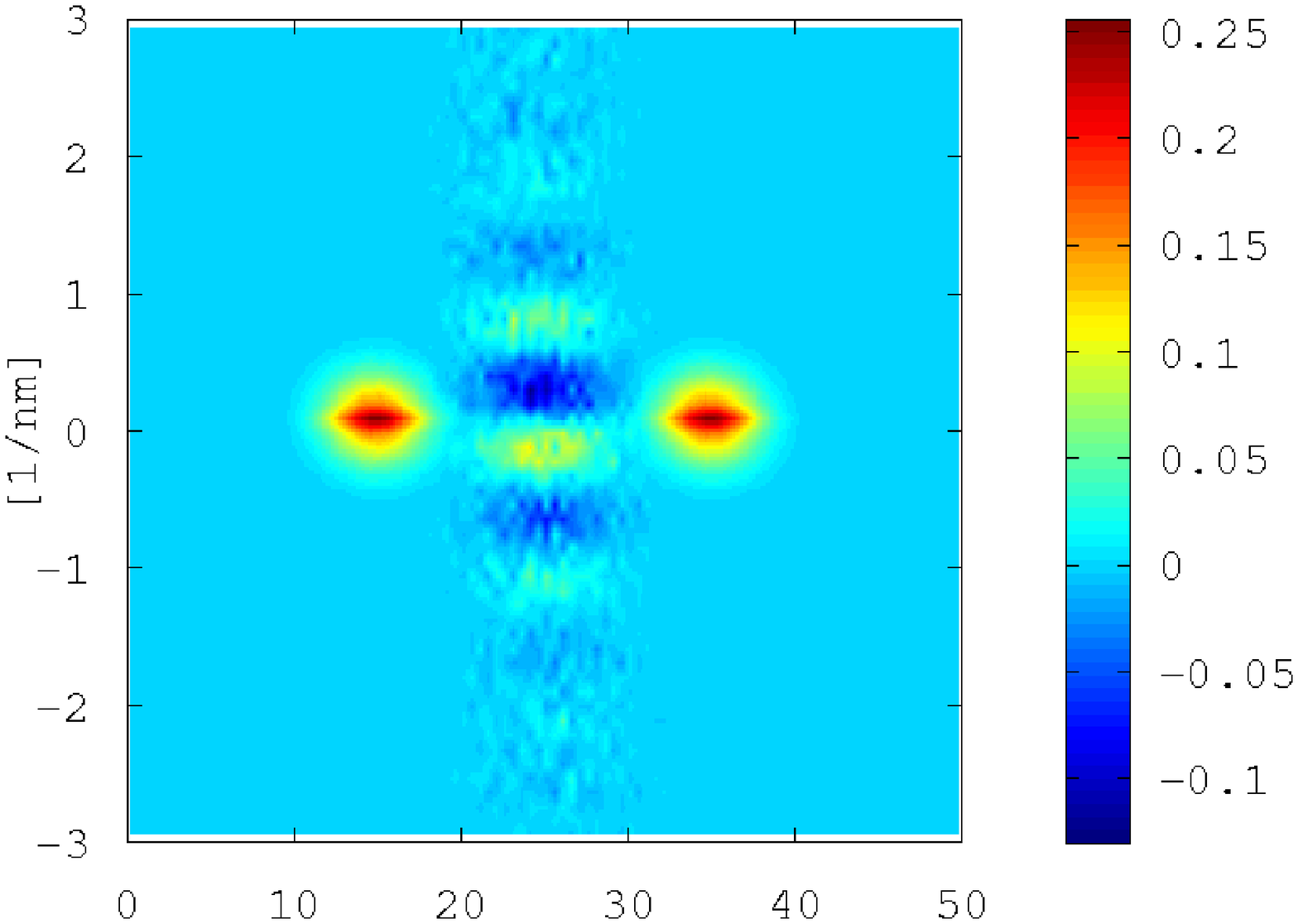}
\\
\includegraphics[width=0.5\textwidth]{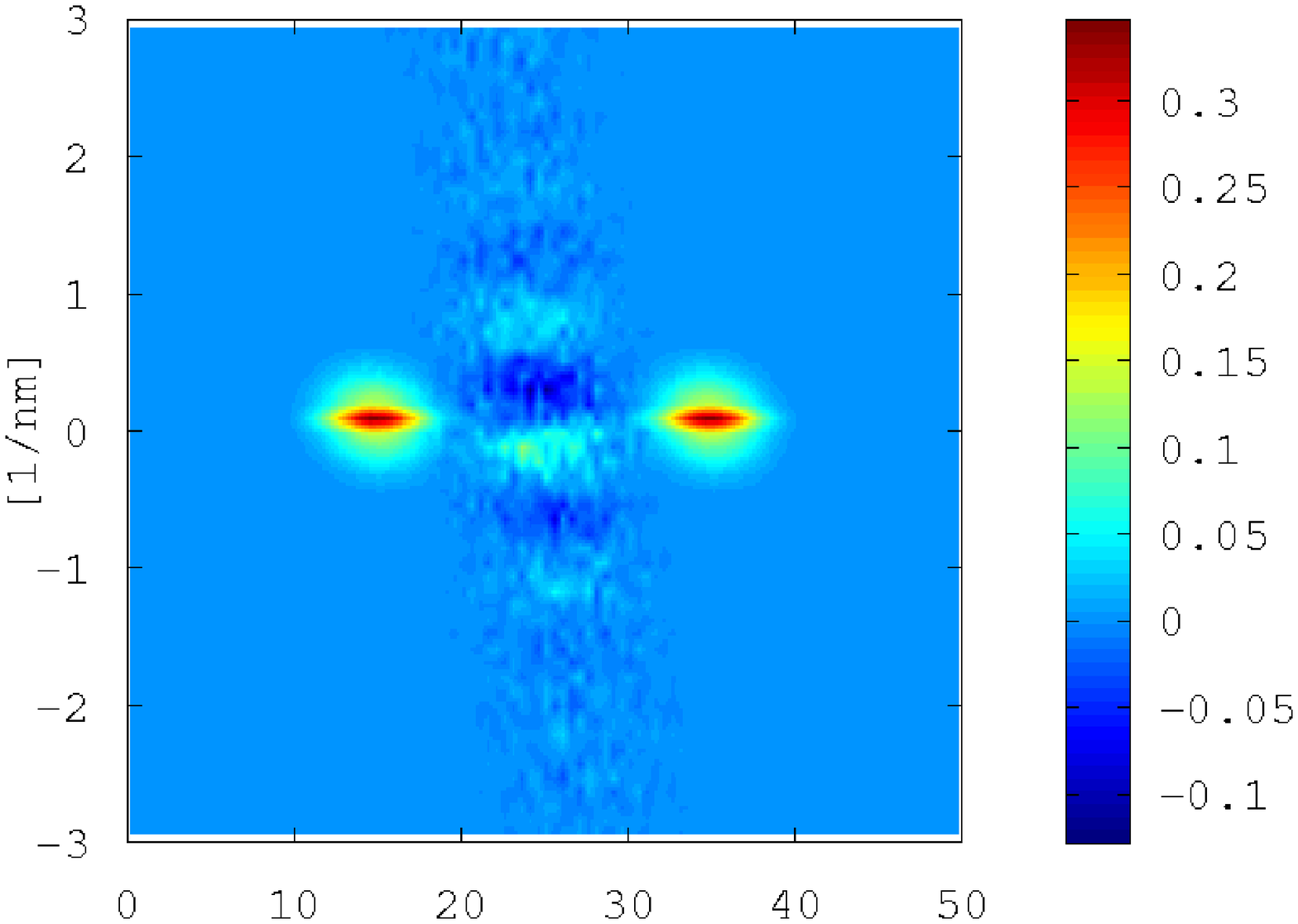}
\includegraphics[width=0.5\textwidth]{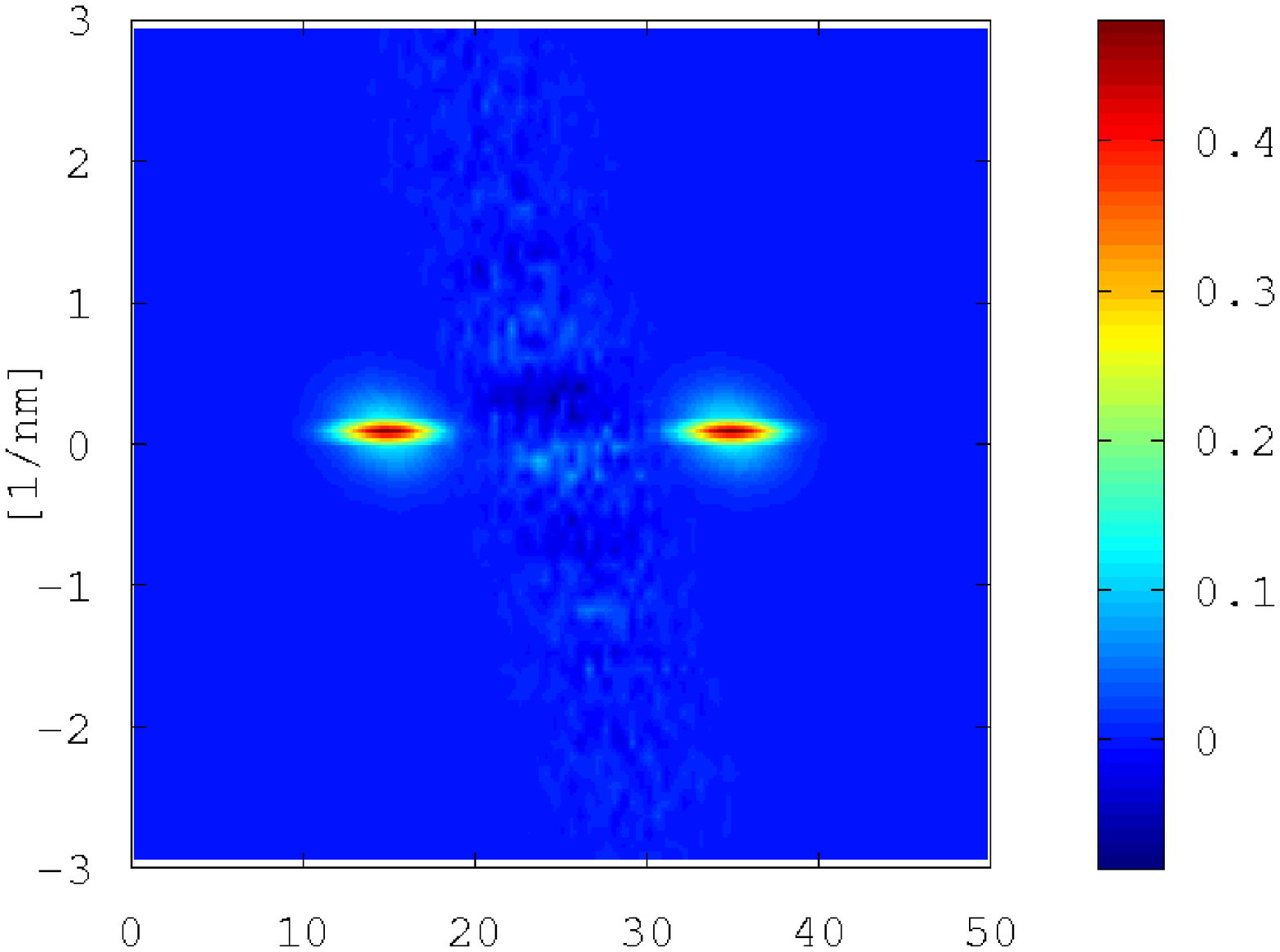}
\end{tabular}
\end{minipage}
\caption{Evolution of an entangled two-body system in the presence of a dissipative background ($1$\% probability of scattering, removing only $5$\% of the energy from a particle, $\sigma_{ent}^x = 3.5$ nm, $\sigma^p_{ent} = \frac{3}{2} \Delta p$). The (normalized) quasi-distribution function in the (reduced) phase-space are shown at times $0.1$ fs, $0.2$ fs, $0.5$ fs and $1$ fs respectively (from top to bottom, left to right). A decay in the entanglement oscillations is, again, observable and a comparison with Fig. \ref{distribution_0.01_0.95_2.5_1.5} suggests that increasing the value $\sigma_{ent}^x$ from $2.5$ nm to $3.5$ nm helps in stabilizing the entanglement between particles.}
\label{distribution_0.01_0.95_3.5_1.5}
\end{figure}

\begin{figure}[h!]
\centering
\begin{minipage}{1.0\textwidth}
\centering
\begin{tabular}{c}
\includegraphics[width=0.5\textwidth]{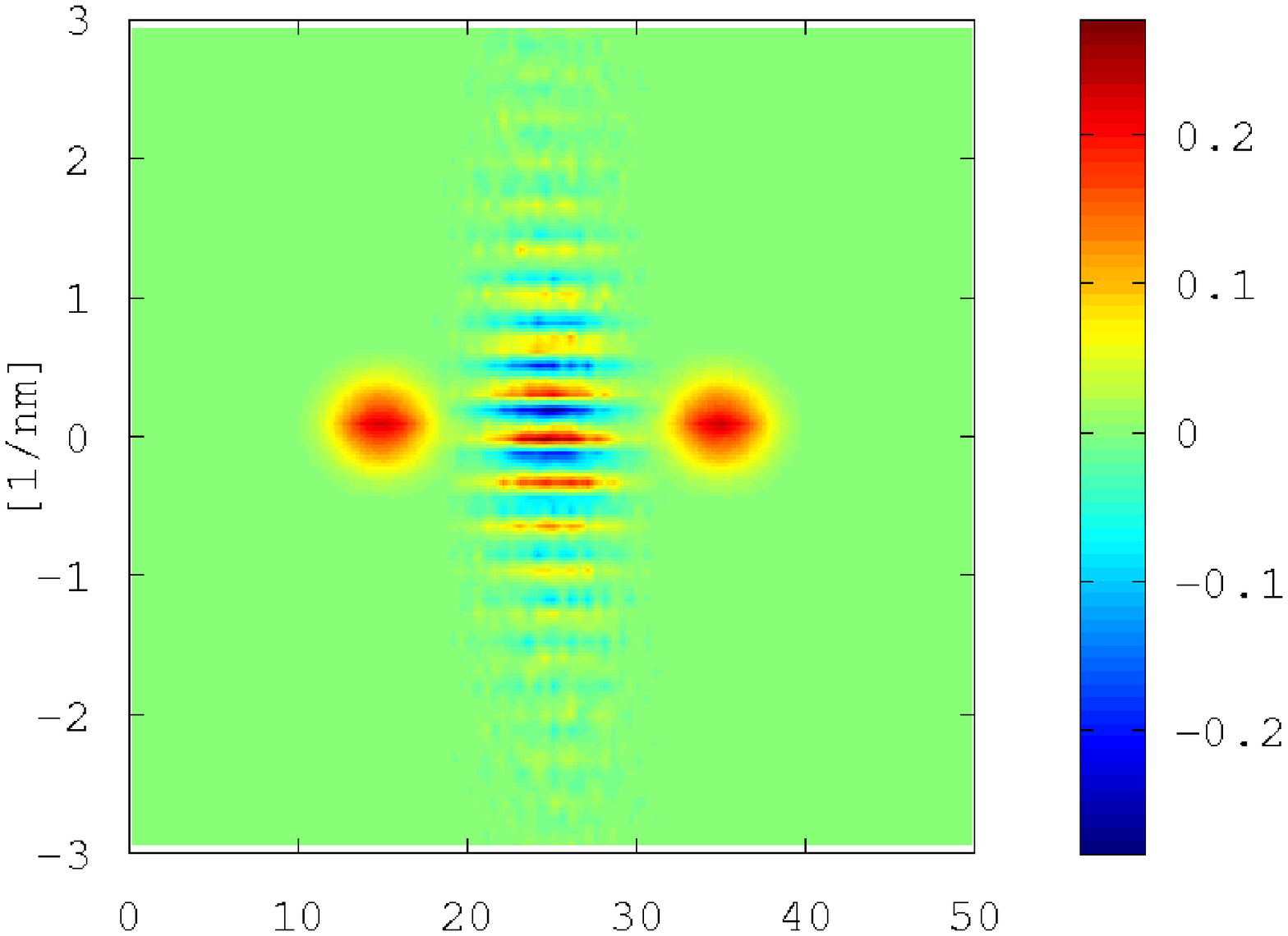}
\includegraphics[width=0.5\textwidth]{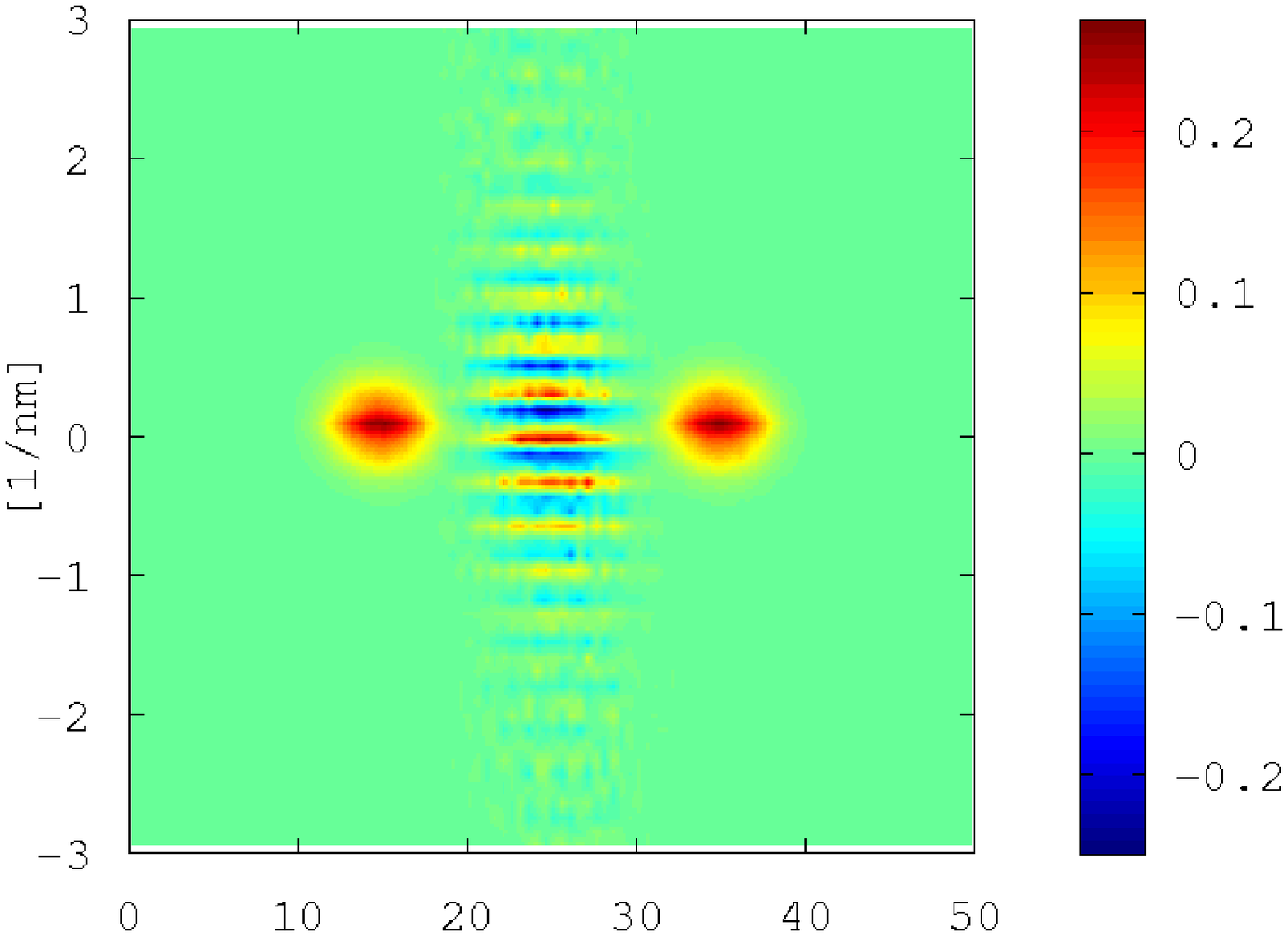}
\\
\includegraphics[width=0.5\textwidth]{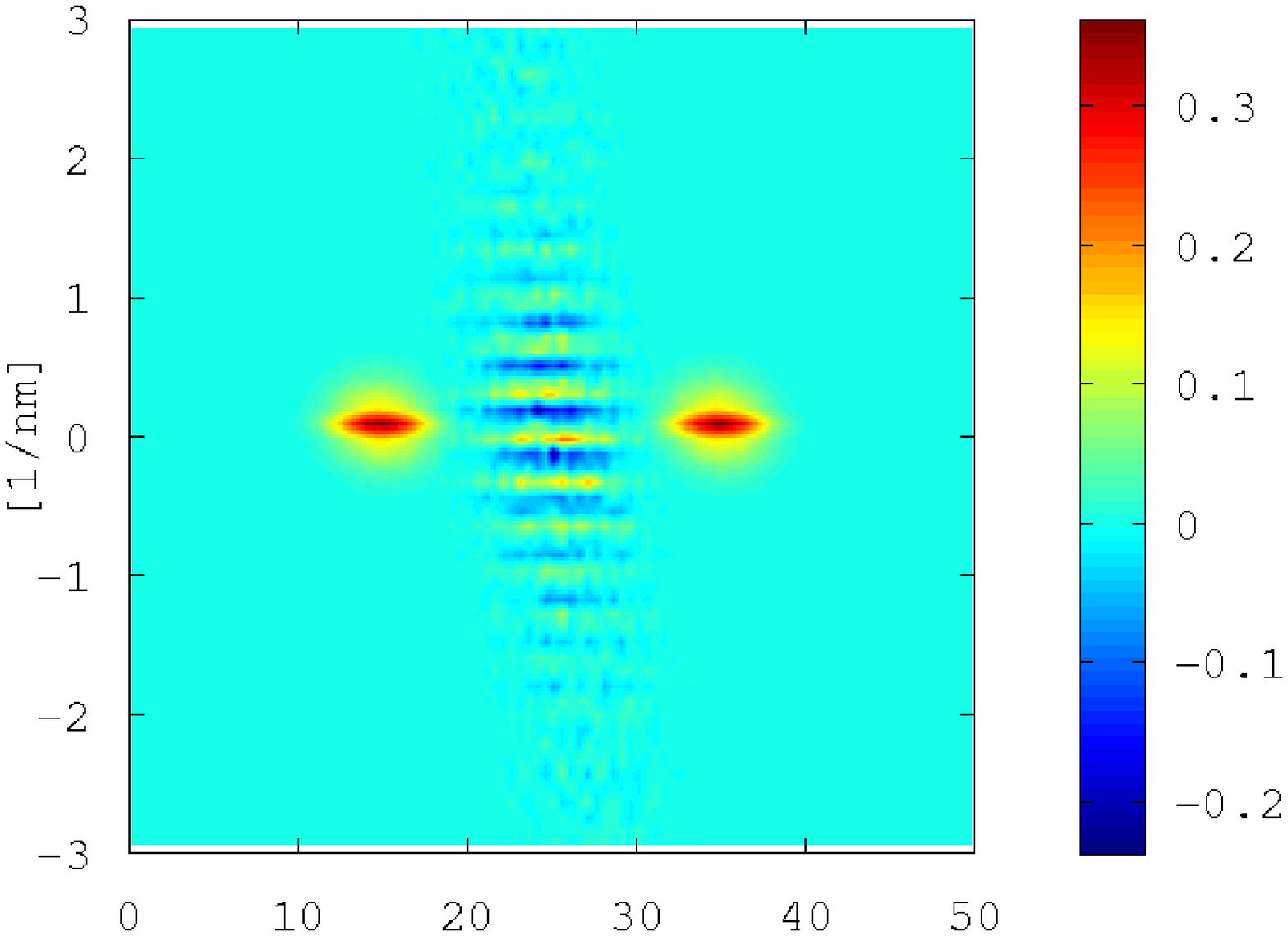}
\includegraphics[width=0.5\textwidth]{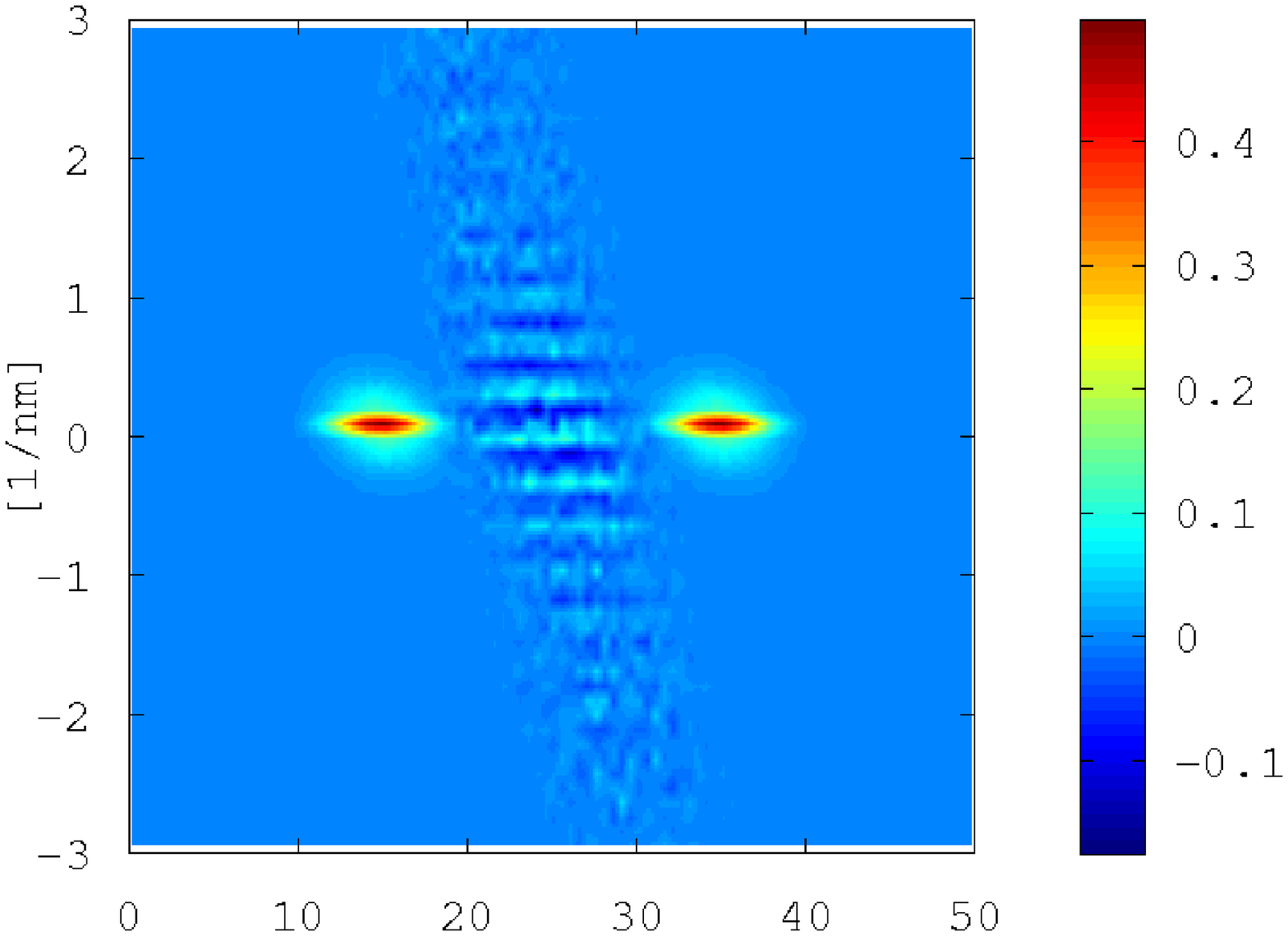}
\end{tabular}
\end{minipage}
\caption{Evolution of an entangled two-body system in the presence of a dissipative background ($1$\% probability of scattering, removing only $5$\% of the energy from a particle, $\sigma_{ent}^x = 3.5$ nm, $\sigma^p_{ent} = \frac{\Delta p}{2}$). The (normalized) quasi-distribution function in the (reduced) phase-space are shown at times $0.1$ fs, $0.2$ fs, $0.5$ fs and $1$ fs respectively (from top to bottom, left to right). The entanglement oscillations decays and, again, a comparison with Fig. \ref{distribution_0.01_0.95_2.5_1.5} suggests that decreasing the value $\sigma_{ent}^p$ stabilizes the entanglement between particles.}
\label{distribution_0.01_0.95_3.5_0.5}
\end{figure}

\begin{figure}[h!]
\centering
\begin{minipage}{1.0\textwidth}
\centering
\begin{tabular}{c}
\includegraphics[width=0.5\textwidth]{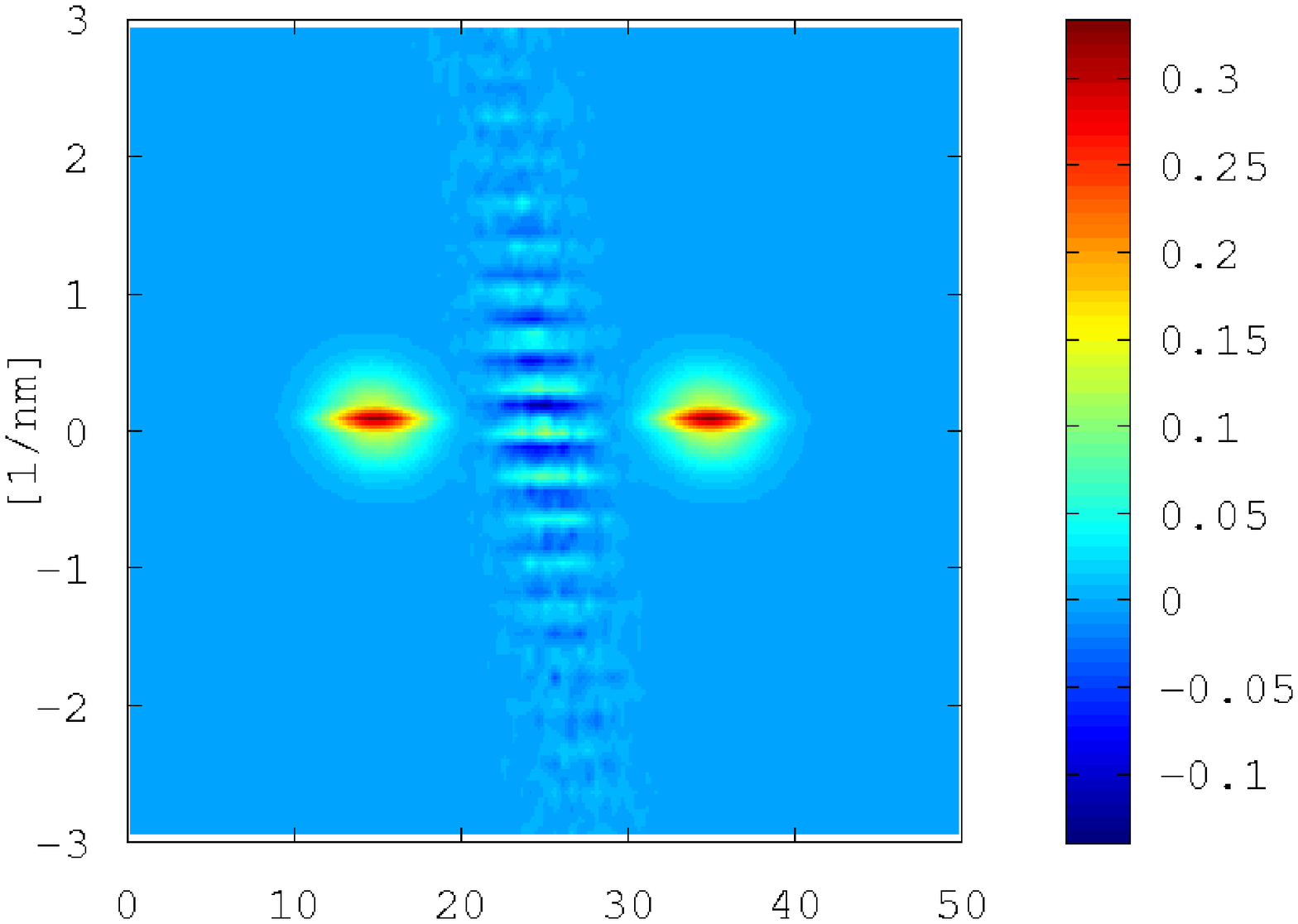}
\\
\includegraphics[width=0.5\textwidth]{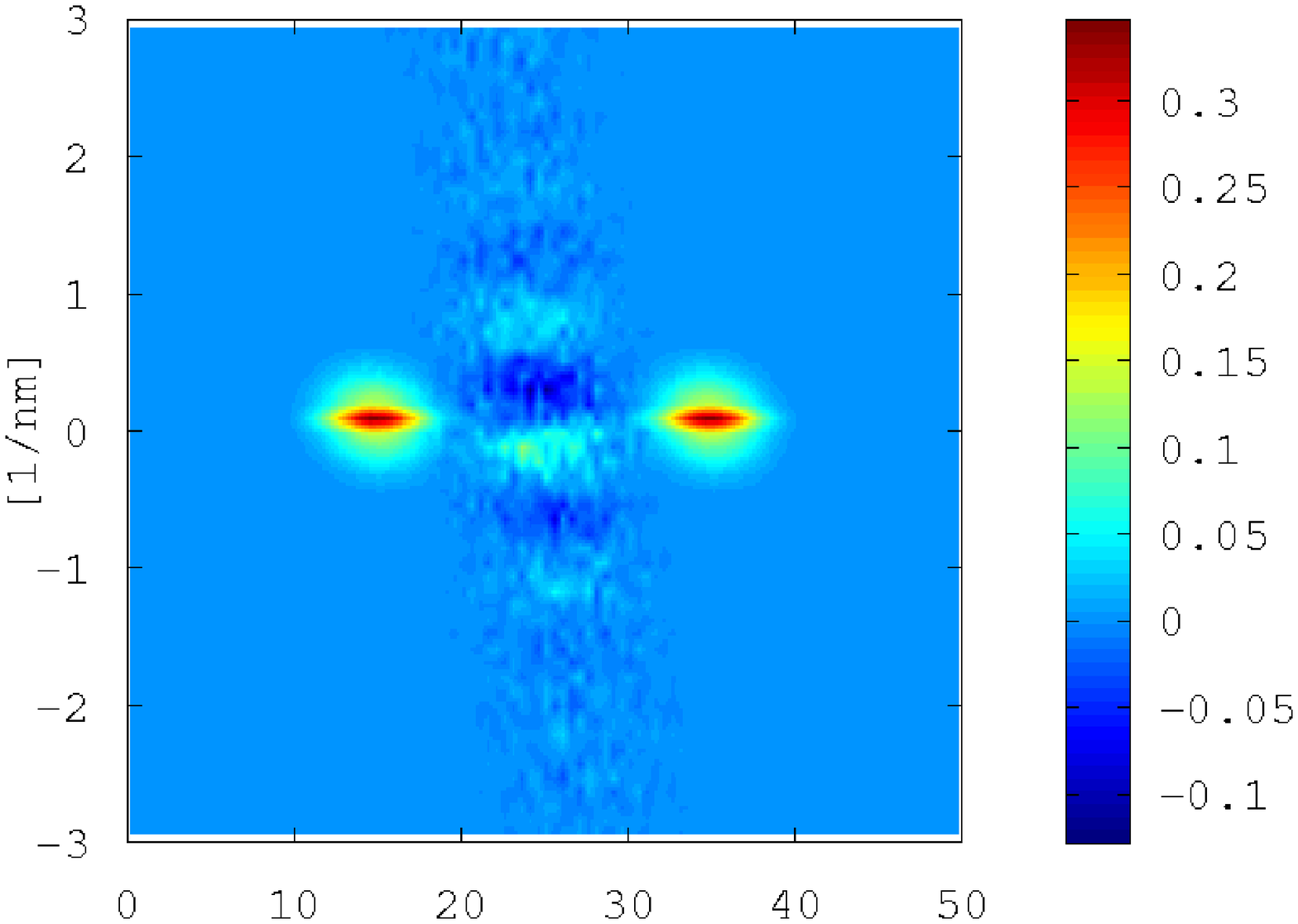}
\\
\includegraphics[width=0.5\textwidth]{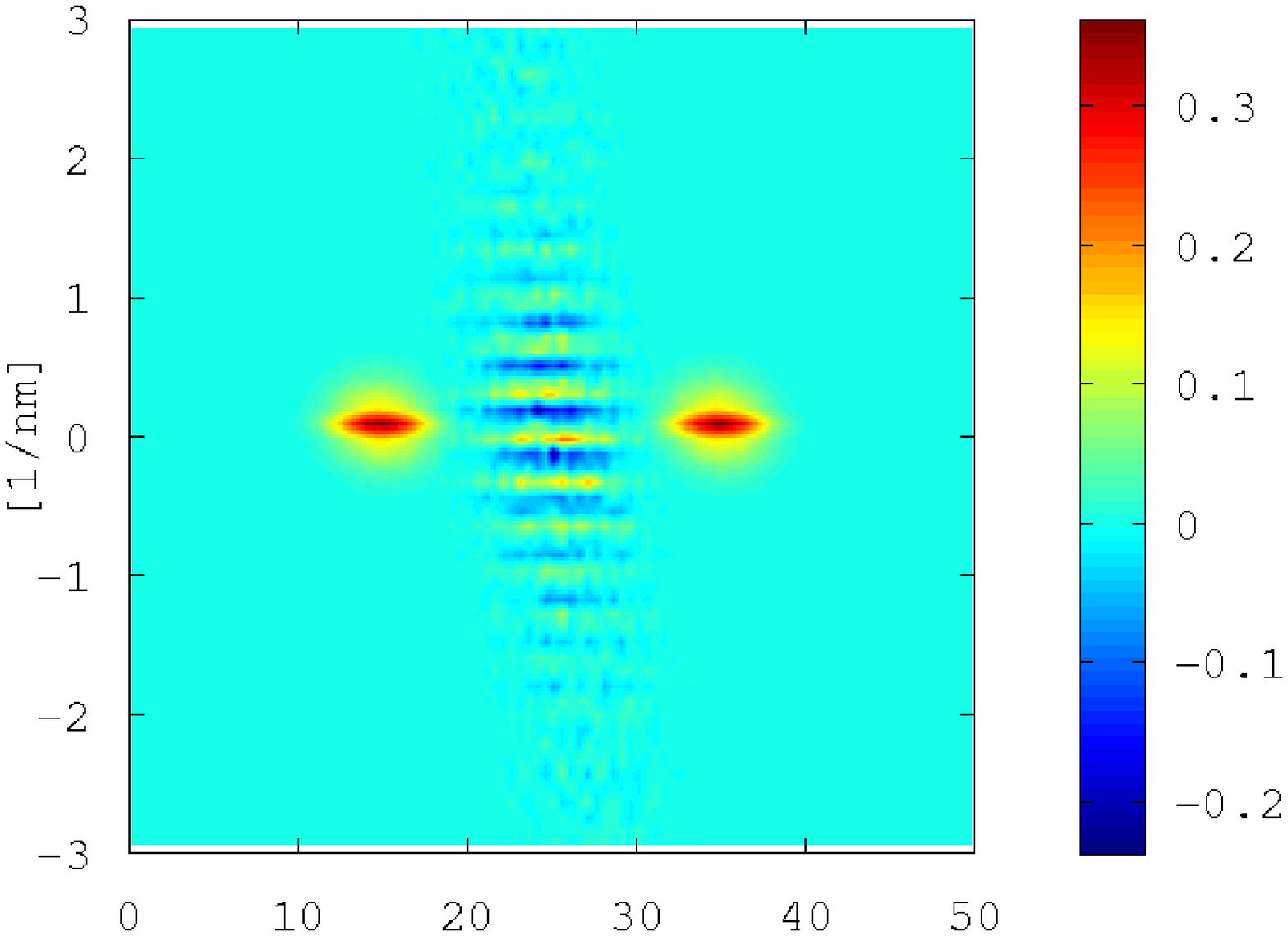}
\end{tabular}
\end{minipage}
\caption{Evolution of the quasi-distribution function at time $0.5$ fs for the cases (top) $\sigma^x_{ent} = 2.5$ nm and $\sigma^p_{ent} = \frac{\Delta p}{2}$, (middle) $\sigma^x_{ent} = 3.5$ nm and $\sigma^p_{ent} = \frac{3}{2} \Delta p$, (bottom) $\sigma^x_{ent} = 3.5$ nm and $\sigma^p_{ent} = \frac{\Delta p}{2} $ respectively.}
\label{distribution_0.5fs}
\end{figure}

The results for different values of $r_{prob}$, $r_{\%}$, $\sigma^x_{ent}$ and $\sigma^p_{ent}$ 
for the normalized quasi-distribution function in a {\sl{reduced}} phase-space
are shown in Figs. \ref{ballistic_2.5_1.5} - \ref{distribution_0.5fs}. In more detail,
Fig. \ref{ballistic_2.5_1.5} represents the ballistic case, i.e. in the total absence
of dissipation ($r_{\%}=0$), used for the purpose of comparison (the values for the
parameters $\sigma^x_{ent}$ and $\sigma^p_{ent}$ are the same as in Fig. \ref{distribution_0.02_0.85_2.5_1.5}).
Significantly, albeit the very long final time (the figure shows a final time equal to $3$ fs,
a very big final time in this particular context), the oscillations due to the entanglement of the
system remain stable. Indeed, the only observable evolution consists of the rotation in the
phase-space of the initial conditions (which is exactly what one expects from a reliable quantum
phase-space model). In distinction, Fig. \ref{distribution_0.02_0.85_2.5_1.5} shows
the very same initial conditions as in Fig. \ref{ballistic_2.5_1.5}, but with the
presence of a noisy background with $r_{prob} = 0.02$ and $r_{\%} = 0.15$. The effects
of dissipation on the entanglement oscillations are immediately apparent. First,
the destruction of the entanglement is quite rapid (importantly at $0.5$ fs it is completely removed).
Second, the reader should note that the two wave-packets are reduced to almost classical
objects (as their dispersion in the momentum space is drastically reduced). As a consequence,
one may consider this as a facilitating the phenomenon of {\sl{quantum decoherence}}.

Therefore, a compelling investigation will be towards an understanding whether entanglement can be
maintained if the noise, or dissipation, in the background is reduced to $r_{prob} = 0.01$
and $r_{\%} = 0.05$ (see Fig. \ref{distribution_0.01_0.95_2.5_1.5} for the results). 
The reader could observe the discernible difference in the behavior between Fig. \ref{distribution_0.02_0.85_2.5_1.5}
and Fig. \ref{distribution_0.01_0.95_2.5_1.5}. In fact, whereas there is a presence of
entanglement oscillations at time $0.5$ fs, at longer times it disappears (see the solution
at $1$ fs at the bottom right of the figure). As ascertained by these preliminary considerations,
the noisy background has an effect on the entanglement of a system, with the increase of noise
leading to faster destruction of the entanglement. 

It is worth examining how the parameters $\sigma^x_{ent}$ and $\sigma^p_{ent}$ in formula (\ref{entangled})
can influence the reinforcement of entanglement. Accordingly, we carry on additional set of
numerical experiments, where the parameters which define the noisy background are kept constant
(i.e. $r_{prob} = 0.01$, $r_{\%} = 0.05$) while $\sigma^x_{ent}$ and $\sigma^p_{ent}$ are varied,
thereupon defining three different cases. Resultantly, Fig. \ref{initial_distributions} represents
the initial conditions corresponding to the values (top) $\sigma^x_{ent}= 2.5$ nm, $\sigma^p_{ent}= \Delta p/2$,
(middle) $\sigma^x_{ent} = 3.5$ nm, $\sigma^p_{ent} = 3/2 \Delta p$ nm$^{-1}$,
(bottom) $\sigma^x_{ent} = 3.5$ nm, $\sigma^p_{ent} = \Delta p/2$ respectively, and 
Figs. \ref{distribution_0.01_0.95_2.5_0.5}, \ref{distribution_0.01_0.95_3.5_1.5}, \ref{distribution_0.01_0.95_3.5_0.5}
depict the corresponding simulation outcomes for the provided initial conditions
(at times $0.1$ fs, $0.2$ fs, $0.5$ fs and $1$ fs respectively).

By comparing Fig. \ref{distribution_0.01_0.95_2.5_0.5} with the Fig. \ref{distribution_0.01_0.95_2.5_1.5}
(same quantity of noise but different values for $\sigma^p_{ent} = \frac{3}{2} \Delta p$), 
it is possible to infer that increasing the number of oscillations in the $p$-direction (and consequently reducing the extension of
every single oscillation) improves the stability in time of the entanglement. As a matter of fact, it is
clear from these figures that, in this particular case, the oscillations can survive for a longer time in spite of the
presence of a dissipative background noise. On the other hand, Fig. \ref{distribution_0.01_0.95_3.5_1.5}
shows the outcome if instead of varying the parameter $\sigma^p_{ent}$ one varies the parameter $\sigma^x_{ent}$
(therefore increasing the size of the oscillations in the $x$-direction).
Thereupon, by comparison with Fig. \ref{distribution_0.01_0.95_2.5_1.5}, it is instantenously recognizable that
this leads to improvement of the resilience of the entanglement, although not as stable as the one obtained
previously (i.e. Fig. \ref{distribution_0.01_0.95_2.5_0.5}). Finally, the case for $\sigma^x_{ent} = 3.5$ nm,
$\sigma^p_{ent} = \Delta p/2$ is shown in Fig. \ref{distribution_0.01_0.95_3.5_0.5} which, compared again to
Fig. \ref{distribution_0.01_0.95_2.5_1.5} displays a better stability of the entanglement in time.
Indeed, the oscillations are still perfectly visible even at the (relatively long) time $t=1$ fs.
To conclude, for the sake of completeness and clarity, a comparison between these three cases at time $0.5$ fs is
reported in Fig. \ref{distribution_0.5fs}.

\section*{\sffamily \Large Conclusions}

In this paper, we extended the recently proposed signed particle formulation of quantum mechanics \cite{SPF}
to the case of many-body problems. This peculiar and novel picture offers advantages rarely available
in other formalisms. It is embarrassingly parallelizable, intuitive and easy to implement.
We have shown that, when restricted to a spatial finite domain and a discrete momentum space,
the proposed extended theory is equivalent to the time-dependent many-body Wigner Monte Carlo method
which has been thoroughly validated in the recent past \cite{PhysRep}.
Interestingly enough, in the context of signed particles the treatment of entangled systems comes naturally.
As a matter of fact, in this work we have been able to apply the formalism to the study of quantum entangled two-body systems
in the presence of a random (but controlled) noisy background, a very difficult (if not daunting) problem
in other formulations of quantum mechanics. We have also shown how the new theory allows the investigation on the resilience
of different types of entanglement.
We believe that such peculiar theory can open the way to tackle problems which are practically impossible in other formalisms
and we encourage the reader to study, run and modify the code available at \cite{nano-archimedes}.
Certainly, the approach utilized in this study to introduce dissipation in the entangled system is rather simple.
An extension to the case of the Wigner-Boltzmann equation (i.e. more realistic scattering effects) in the two-body context will be the subject of a
further future work. This could provide an even better comprehension of the entanglement phenomenon in technology relevant situations.

\subsection*{\sffamily \large ACKNOWLEDGMENTS}

J.M.~Sellier would like to thank M. Anti for her support and encouragement.
K.G.~Kapanova has been partially supported by the Bulgarian Science Fund under grant DFNI 02/20.

\clearpage

%%%%%%%%%%%%%%%%%%%%%%%%%%%%%%%%%%%%%%%%%%%%%%%%%%%%%%%%%%%%%%%%%%%%%%%%%%%%%%%%%
% BIBLIOGRAPHY

%\bibliography{bibtexrefs}   % Produces the bibliography via BibTeX.

%%%%%%%%%%%%%%%%%%%%%%%%%%%%%%%%%%%%%%%%%%%%%%%%%%%%%%%%%%%%%%%%%%%%%%%%%%%%%%%%%

\clearpage
%%%%%%%%%%%%%%%%%%%%%%%%%%%%%%%%%%%%%%%%%%%%%%%%%%%%%%%%%%%%%%%%%%%%%%%%%%%%%%%%%
% FIGURE CAPTIONS

%\begin{figure}[h!]
%\centering
%\begin{minipage}{1.0\textwidth}
%\centering
%\begin{tabular}{c}
%\includegraphics[width=0.95\textwidth]{distribution_0.01_0.95_3.5nm_1.5DKX_0fs.eps}
%\end{tabular}
%\end{minipage}
%\caption{Initial conditions of an entangled two-body system with $1$\% probability of scattering, removing only $5$\% of the energy from a particle, $\sigma^{ent}_x = 3.5$ nm, and $\sigma_k^{ent} = \frac{3}{2} \Delta_k$.}
%\label{initial_distribution_0.01_0.95_3.5_1.5}
%\end{figure}

%%%%%%%%%%%%%%%%%%%%%%%%%%%%%%%%%%%%%%%%%%%%%%%%%%%%%%%%%%%%%%%%%%%%%%%%%%%%%%%%%
% FIGURE FILES

\clearpage

\end{document}